\documentclass[aps,prc,twocolumn,showpacs,preprintnumbers,nofootinbib,float,superscriptaddress,longbibliography]{revtex4-2}
\usepackage{graphicx,fancybox}
\usepackage{amsmath,amssymb}
\usepackage[colorlinks=true, pdfstartview=FitV, linkcolor=red, citecolor=blue, urlcolor=blue]{hyperref}
\usepackage{color}
\usepackage{soul}
\usepackage{array}
\usepackage{url}
\usepackage[utf8]{inputenc}
\usepackage{lineno}
\newcommand*\patchAmsMathEnvironmentForLineno[1]{%
  \expandafter\let\csname old#1\expandafter\endcsname\csname #1\endcsname
  \expandafter\let\csname oldend#1\expandafter\endcsname\csname end#1\endcsname
  \renewenvironment{#1}%
     {\linenomath\csname old#1\endcsname}%
     {\csname oldend#1\endcsname\endlinenomath}}%
\newcommand*\patchBothAmsMathEnvironmentsForLineno[1]{%
  \patchAmsMathEnvironmentForLineno{#1}%
  \patchAmsMathEnvironmentForLineno{#1*}}%
\AtBeginDocument{%
\patchBothAmsMathEnvironmentsForLineno{equation}%
\patchBothAmsMathEnvironmentsForLineno{align}%
\patchBothAmsMathEnvironmentsForLineno{flalign}%
\patchBothAmsMathEnvironmentsForLineno{alignat}%
\patchBothAmsMathEnvironmentsForLineno{gather}%
\patchBothAmsMathEnvironmentsForLineno{multline}%

}
\usepackage[normalem]{ulem}

\newcommand{\pythia}{\textsc{pythia}}

\newcounter{get_page_number}
\def\pagelabel#1{\refstepcounter{get_page_number}\label{#1}}

\usepackage{catchfilebetweentags}

\makeatletter
\newcommand*{\newreptext}[1]{%
  \begingroup 
    \csname @safe@actives@true\endcsname
  \expandafter\endgroup
  \expandafter\newcommand\csname reptext@#1\endcsname
}

\newcommand*{\reptext}[1]{%
  \begingroup
  \csname @safe@actives@true\endcsname 
  \@ifundefined{reptext@#1}{%
    \@latex@error{\string\reptext{#1} is undefined}\@ehc
    \endgroup
    \textbf{??}%
  }{%
    \endgroup
    \@nameuse{reptext@#1}%
  }%
}
\makeatother

\begin{document}

\title{Hard jet substructure in a multistage approach}

\newcommand{\AiuAddress}{Akita International University, Yuwa, Akita-city 010-1292, Japan.}
\newcommand{\McGillAddress}{Department of Physics, McGill University, Montr\'{e}al, Qu\'{e}bec H3A\,2T8, Canada.}
\newcommand{\WaynePhysAddress}{Department of Physics and Astronomy, Wayne State University, Detroit, Michigan 48201, USA.}
\newcommand{\ReginaAddress}{Department of Physics, University of Regina, Regina, Saskatchewan S4S 0A2, Canada.}
\newcommand{\LLNLAddress}{Lawrence Livermore National Laboratory, Livermore, California 94550, USA.}
\newcommand{\TexasAMCompAddress}{Research Computing Group, University Technology Solutions, The University of Texas at San Antonio, San Antonio, Texas 78249, USA.}
\newcommand{\DukePhysAddress}{Department of Physics, Duke University, Durham, North Carolina 27708, USA}
\newcommand{\ShandongAddress}{Institute of Frontier and Interdisciplinary Science, Shandong University, Qingdao, Shandong 266237, China}
\newcommand{\MITNuclAddress}{Laboratory for Nuclear Science, Massachusetts Institute of Technology, Cambridge, Massachusetts 02139, USA.}
\newcommand{\MITPhysAddress}{Department of Physics, Massachusetts Institute of Technology, Cambridge, Massachusetts 02139, USA.}
\newcommand{\VandyAddress}{Department of Physics and Astronomy, Vanderbilt University, Nashville, Tennessee 37235, USA.}
\newcommand{\UTNAddress}{Department of Physics and Astronomy, University of Tennessee, Knoxville, Tennessee 37996, USA.}
\newcommand{\OrkRidgeAddress}{Physics Division, Oak Ridge National Laboratory, Oak Ridge, Tennessee 37830, USA.}
\newcommand{\UCBerkeleyAddress}{Department of Physics, University of California, Berkeley, California 94270, USA.}
\newcommand{\LBLAddress}{Nuclear Science Division, Lawrence Berkeley National Laboratory, Berkeley, California 94270, USA.}
\newcommand{\GSIAddress}{GSI Helmholtzzentrum f\"{u}r Schwerionenforschung, 64291 Darmstadt, Germany.}
\newcommand{\GoetheAddress}{Institute for Theoretical Physics, Goethe University, 60438 Frankfurt am Main, Germany.}
\newcommand{\FIASAddress}{Frankfurt Institute for Advanced Studies, 60438 Frankfurt am Main, Germany.}
\newcommand{\TexasAMCyclAddress}{Cyclotron Institute, Texas A\&M University, College Station, Texas 77843, USA.}
\newcommand{\TexasAMPhysAddress}{Department of Physics and Astronomy, Texas A\&M University, College Station, Texas 77843, USA.}
\newcommand{\SCNUKeyLabAddress}{Guangdong Provincial Key Laboratory of Nuclear Science, Institute of Quantum Matter, South China Normal University, Guangzhou 510006, China.}
\newcommand{\SCNUJointLabAddress}{Guangdong-Hong Kong Joint Laboratory of Quantum Matter, Southern Nuclear Science Computing Center, South China Normal University, Guangzhou 510006, China.}
\newcommand{\OhioAddress}{Department of Physics, The Ohio State University, Columbus, Ohio 43210, USA.}
\newcommand{\DukeStatAddress}{Department of Statistical Science, Duke University, Durham, North Carolina 27708, USA.}
\newcommand{\CCNUKeyAndIOPPAddress}{Key Laboratory of Quark and Lepton Physics (MOE) and Institute of Particle Physics, Central China Normal University, Wuhan 430079, China.}
\newcommand{\BNLAddress}{Department of Physics, Brookhaven National Laboratory, Upton, New York 11973, USA.}
\newcommand{\LANLAddress}{Los Alamos National Laboratory, Theoretical Division, Los Alamos, New Mexico 87545, USA.}
\newcommand{\WayneCompAddress}{Department of Computer Science, Wayne State University, Detroit, Michigan 48202, USA.}
\newcommand{\INTAddress}{Institute for Nuclear Theory, University of Washington, Seattle, Washington 98195, USA.}
\newcommand{\SaoPauloAddress}{Instituto de F\`{i}sica, Universidade de S\~{a}o Paulo, C.P. 66318, 05315-970 S\~{a}o Paulo, S\~{a}o Paulo, Brazil.}
\newcommand{\RIKENBNLAddress}{RIKEN BNL Research Center, Brookhaven National Laboratory, Upton, New York 11973, USA.}
\newcommand{\KentAddress}{Department of Physics, Kent State University, Kent, Ohio 44242, USA.}

\author{Y.~Tachibana}
\email[Corresponding author: ]{ytachibana@aiu.ac.jp}
\affiliation{\AiuAddress}

\author{A.~Kumar}
\email[Corresponding author: ]{amit.kumar@uregina.ca}
\affiliation{\McGillAddress}
\affiliation{\WaynePhysAddress}
\affiliation{\ReginaAddress}

\author{A.~Majumder}
\affiliation{\WaynePhysAddress}

\author{A.~Angerami}
\affiliation{\LLNLAddress}

\author{R.~Arora}
\affiliation{\TexasAMCompAddress}

\author{S.~A.~Bass}
\affiliation{\DukePhysAddress}

\author{S.~Cao}
\affiliation{\ShandongAddress}
\affiliation{\WaynePhysAddress}

\author{Y.~Chen}
\affiliation{\MITNuclAddress}
\affiliation{\MITPhysAddress}
\affiliation{\VandyAddress}

\author{T.~Dai}
\affiliation{\DukePhysAddress}

\author{L.~Du}
\affiliation{\McGillAddress}

\author{R.~Ehlers}
\affiliation{\UTNAddress}
\affiliation{\OrkRidgeAddress}
\affiliation{\UCBerkeleyAddress}
\affiliation{\LBLAddress}

\author{H.~Elfner}
\affiliation{\GSIAddress}
\affiliation{\GoetheAddress}
\affiliation{\FIASAddress}

\author{W.~Fan}
\affiliation{\DukePhysAddress}

\author{R.~J.~Fries}
\affiliation{\TexasAMCyclAddress}
\affiliation{\TexasAMPhysAddress}

\author{C.~Gale}
\affiliation{\McGillAddress}

\author{Y.~He}
\affiliation{\SCNUKeyLabAddress}
\affiliation{\SCNUJointLabAddress}

\author{M.~Heffernan}
\affiliation{\McGillAddress}

\author{U.~Heinz}
\affiliation{\OhioAddress}

\author{B.~V.~Jacak}
\affiliation{\UCBerkeleyAddress}
\affiliation{\LBLAddress}

\author{P.~M.~Jacobs}
\affiliation{\UCBerkeleyAddress}
\affiliation{\LBLAddress}

\author{S.~Jeon}
\affiliation{\McGillAddress}

\author{Y.~Ji}
\affiliation{\DukeStatAddress}

\author{K.~Kauder}
\affiliation{\BNLAddress}

\author{L.~Kasper}
\affiliation{\VandyAddress}

\author{W.~Ke}
\affiliation{\LANLAddress}

\author{M.~Kelsey}
\affiliation{\WaynePhysAddress}

\author{M.~Kordell~II}
\affiliation{\TexasAMCyclAddress}
\affiliation{\TexasAMPhysAddress}

\author{J.~Latessa}
\affiliation{\WayneCompAddress}

\author{Y.-J.~Lee}
\affiliation{\MITNuclAddress}
\affiliation{\MITPhysAddress}

\author{D.~Liyanage}
\affiliation{\OhioAddress}

\author{A.~Lopez}
\affiliation{\SaoPauloAddress}

\author{M.~Luzum}
\affiliation{\SaoPauloAddress}

\author{S.~Mak}
\affiliation{\DukeStatAddress}

\author{A.~Mankolli}
\affiliation{\VandyAddress}

\author{C.~Martin}
\affiliation{\UTNAddress}

\author{H.~Mehryar}
\affiliation{\WayneCompAddress}

\author{T.~Mengel}
\affiliation{\UTNAddress}

\author{J.~Mulligan}
\affiliation{\UCBerkeleyAddress}
\affiliation{\LBLAddress}

\author{C.~Nattrass}
\affiliation{\UTNAddress}

\author{D.~Oliinychenko}
\affiliation{\LBLAddress}
\affiliation{\INTAddress}

\author{J.-F. Paquet}
\affiliation{\DukePhysAddress}
\affiliation{\VandyAddress} 

\author{J.~H.~Putschke}
\affiliation{\WaynePhysAddress}

\author{G.~Roland}
\affiliation{\MITNuclAddress}
\affiliation{\MITPhysAddress}

\author{B.~Schenke}
\affiliation{\BNLAddress}

\author{L.~Schwiebert}
\affiliation{\WayneCompAddress}

\author{A.~Sengupta}
\affiliation{\TexasAMCyclAddress}
\affiliation{\TexasAMPhysAddress}

\author{C.~Shen}
\affiliation{\WaynePhysAddress}
\affiliation{\RIKENBNLAddress}

\author{A.~Silva}
\affiliation{\UTNAddress}

\author{C.~Sirimanna}
\affiliation{\WaynePhysAddress}
\affiliation{\DukePhysAddress}

\author{D.~Soeder}
\affiliation{\DukePhysAddress}

\author{R.~A.~Soltz}
\affiliation{\WaynePhysAddress}
\affiliation{\LLNLAddress}

\author{I.~Soudi}
\affiliation{\WaynePhysAddress}

\author{J.~Staudenmaier}
\affiliation{\GoetheAddress}

\author{M.~Strickland}
\affiliation{\KentAddress}

\author{J.~Velkovska}
\affiliation{\VandyAddress}

\author{G.~Vujanovic}
\affiliation{\ReginaAddress}

\author{X.-N.~Wang}
\affiliation{\CCNUKeyAndIOPPAddress}
\affiliation{\UCBerkeleyAddress}
\affiliation{\LBLAddress}

\author{R.~L.~Wolpert}
\affiliation{\DukeStatAddress}

\author{W.~Zhao}
\affiliation{\WaynePhysAddress}
\affiliation{\UCBerkeleyAddress}
\affiliation{\LBLAddress}

\collaboration{JETSCAPE Collaboration}

\begin{abstract}
We present predictions and postdictions for a wide variety of hard jet-substructure observables using a multistage model within the \textsc{jetscape} framework. 
The details of the multistage model and the various parameter choices are described in [\href{https://journals.aps.org/prc/abstract/10.1103/PhysRevC.107.034911}{Phys. Rev. C 107, 034911 (2023)}]. 
A novel feature of this model is the presence of two stages of jet modification: a high virtuality phase [modeled using the \textsc{matter} modular all
twist transverse-scattering elastic-drag and radiation model (\textsc{matter})], where modified coherence effects diminish medium-induced radiation, and a lower virtuality phase [modeled using the linear Boltzmann transport model (\textsc{lbt})], where parton splits are fully resolved by the medium as they endure multiple scattering induced energy loss. Energy-loss calculations are carried out on event-by-event viscous fluid dynamic backgrounds constrained by experimental data. The uniform and consistent descriptions of multiple experimental observables demonstrate the essential role of modified coherence effects and the multistage modeling of jet evolution. 
Using the best choice of parameters from [\href{https://journals.aps.org/prc/abstract/10.1103/PhysRevC.107.034911}{Phys. Rev. C 107, 034911 (2023)}], and with no further tuning, 
we present calculations for the medium modified jet fragmentation function, the groomed jet momentum fraction $z_g$ and angular separation $r_g$ distributions, as well as the nuclear modification factor of groomed jets. These calculations provide accurate descriptions of published data from experiments at the Large Hadron Collider. Furthermore, we provide predictions from the multistage model for future measurements at the BNL Relativistic Heavy Ion Collider. 
\end{abstract}

\maketitle


\section{Introduction}
\label{Section:Intro}
In high-energy heavy-ion collisions, the high-transverse momentum ($p_{T}$) partons ($p_{T} \gtrapprox 10$~GeV) are generated almost at the instant at which the incoming nuclei overlap. Such high $p_{T}$ partons are generated in parton-parton exchanges with large momentum transfers $Q \gg \Lambda_{\mathrm{QCD}}$. They are typically produced far from their mass shell and engender multiple collinear emissions produced over a large time range. In the case of a heavy-ion collision, the propagation and development of these parton showers are strongly affected by the produced Quark Gluon Plasma (QGP).
Studying jet modification in nucleus-nucleus collisions relative to proton-proton collisions, together with constraints from model-to-data comparison provides unique opportunities to probe the properties of the QGP~\cite{Bjorken:1982tu:Manual,Appel:1985dq,Baier:1996kr,Baier:1996sk,Zakharov:1996fv,Gyulassy:1999zd,Gyulassy:2000fs,Gyulassy:2000er,Wiedemann:2000za,Wiedemann:2000tf,Guo:2000nz,Wang:2001ifa,Majumder:2009ge,Arnold:2001ba,Arnold:2002ja,Majumder:2010qh,Blaizot:2015lma,Qin:2015srf,Cao:2020wlm}.

The experimental attempts started at the Relativistic Heavy Ion Collider (RHIC) with the observation of suppression in the yield of single inclusive hadrons~\cite{PHENIX:2001hpc,PHENIX:2003djd,PHENIX:2003qdj,STAR:2002ggv,STAR:2003fka} 
and associated hadrons (dihadrons)~\cite{STAR:2002svs, STAR:2005ryu,PHENIX:2007yjc}
produced with high transverse momentum relative to the yield in proton-proton collisions. 
Since 2010, starting at the Large Hadron Collider (LHC) and later at RHIC, the ability of experiments evolved from single hadrons and dihadrons to jets~\cite{ALICE:2013dpt,ATLAS:2010isq,CMS:2011iwn}. 

Over the last decade, experiments have attained the ability to not just study the energy-momentum and cross section of a jet but also to look at modifications of the internal properties of the jet, often referred to as \emph{jet substructure}. Based on current detector improvements and accumulated high statistics data at RHIC and the LHC, it is possible to analyze a vast variety of observables revealing different aspects of the jet-medium interaction~\cite{Connors:2017ptx}. 
For example, the yield suppression and internal structure of fully reconstructed jets, revealed in observables such as the jet fragmentation function and jet shape (respectively), provide details on the diffusion of jet energy and momentum in momentum or angular space due to the interaction with the medium~\cite{ATLAS:2010isq,CMS:2011iwn,ATLAS:2012tjt,ALICE:2013dpt,ATLAS:2014ipv,CMS:2016uxf,STAR:2016dfv,ATLAS:2018gwx,ALICE:2019qyj,CMS:2021vui,STAR:2020xiv,CMS:2013lhm,CMS:2016cvr,CMS:2018zze,CMS:2018jco,ALICE:2019whv,CMS:2021nhn,CMS:2014jjt,ATLAS:2014dtd,ATLAS:2017nre,ATLAS:2018bvp,ATLAS:2019dsv,ATLAS:2019pid}. 
Even the structural modification of hard partonic branching is now potentially accessible through groomed jet observables~\cite{CMS:2017qlm,Kauder:2017cvz,CMS:2018fof,ALICE:2019ykw,ALargeIonColliderExperiment:2021mqf,ATLAS:2022vii}. 

On the theory side, many studies have attempted to describe and understand the jet-medium interaction by constructing models that reproduce these various observables or propose predictions and new observables~
\cite{Vitev:2009rd,Qin:2010mn,Casalderrey-Solana:2011rbm,He:2011pd,Qin:2012gp,Blaizot:2014ula,Chien:2015hda,Chang:2016gjp,Mehtar-Tani:2016aco,Chen:2016vem,Chien:2016led,Mehtar-Tani:2017web,Chang:2017gkt,Tachibana:2017syd,Li:2018xuv,Chang:2019sae,Qiu:2019sfj,Ringer:2019rfk,Cao:2021rpv,Mehtar-Tani:2021fud,Sirimanna:2022zje}. 
In particular, to obtain a universal understanding, it is essential to simultaneously explain multiple observables, ultimately all observables, with a consistent theoretical picture. 
Therefore, Monte Carlo calculations, which can generate experiment-like events by a single model, are a powerful tool for theoretical approaches because they enable one to calculate a wide range of event-by-event defined jet observables~\cite{Lokhtin:2005px,Zapp:2008gi,Renk:2008xq,Armesto:2009fj,Schenke:2009gb,Li:2010ts,Young:2011qx,Lokhtin:2011qq,Zapp:2013vla,Casalderrey-Solana:2014bpa,He:2015pra,Casalderrey-Solana:2016jvj,Cao:2016gvr,KunnawalkamElayavalli:2017hxo,Milhano:2017nzm,Chen:2017zte,He:2018xjv,Luo:2018pto,Park:2018acg,Ke:2018jem,Casalderrey-Solana:2019ubu,Pablos:2019ngg,Caucal:2019uvr,Ke:2020clc,Dai:2020rlu,Chen:2020tbl,Zhao:2021vmu,Liu:2021dpm,Luo:2021hoo,Luo:2021voy,Caucal:2021cfb,Yazdi:2022bru,Yang:2022nei,Shi:2022rja,Cao:2024pxc}. 

Jets evolve dynamically, moving through the expanding medium, and generating more partons from splits and interactions with the dense medium. 
The original partons start at very high virtuality, and thus, the early splits have a small transverse size. 
These splittings from the leading parton and the still highly-virtual daughters are driven by their individual virtualities, with minor medium correction via the scattering, strongly suppressed due to their small transverse size. We refer to these as Vacuum Like Emissions (VLE)~\cite{Caucal:2018dla}. 
To simulate the VLEs, taking into account the reduction in the effective interaction rate with scale dependence, an event generator such as modular all twist transverse-scattering elastic-drag and radiation (\textsc{matter})~\cite{Cao:2017qpx,Majumder:2013re} can be employed.

With repeated splittings, the virtuality of the partons reduces to the point that splits are widely separated in time.  
With decreasing virtuality, the transverse size of the parton becomes larger, thereby increasing the rate of interaction with the medium, which in turn triggers more radiation. 
Thus, the main mechanism causing parton splittings changes dynamically in the medium. 
The evolution of such partons at lower virtuality but energy still large enough to treat the medium interaction perturbatively can be approximated by kinetic theory-based approaches for on-shell particles, as implemented by generators such as linear Boltzmann transport (\textsc{lbt})~\cite{He:2015pra,Cao:2016gvr,He:2018xjv,Luo:2018pto}, or the modular algorithm for
relativistic treatment of heavy ion interactions (\textsc{martini})~\cite{Schenke:2009gb,Yazdi:2022bru,Shi:2022rja}.  
As partons transition to energies and virtualities close to those of the QGP, they begin to undergo strong coupling~\cite{Casalderrey-Solana:2014bpa} and thermalization with the medium~\cite{Tachibana:2020mtb}. 
Thus, jets interact with the medium over a wide range of scales, which requires incorporating multiple generators at different scales for simulations~\cite{Majumder:2010qh}. 

\textsc{jetscape} is a general-purpose framework for Monte Carlo simulations of the complete evolution in high-energy heavy-ion collisions~\cite{Putschke:2019yrg,JETSCAPE:2019udz,JETSCAPE:2020shq,JETSCAPE:2020mzn,JETSCAPE:2021ehl:Manual,JETSCAPE:2022cob,JETSCAPE:2022jer,JETSCAPE:2022hcb,JETSCAPE:2023ikg}. 
The framework is designed to be as general and extensive as possible while modularizing each physics element involved in a collision event, such as the generation of geometric initial conditions, hydrodynamic evolution of the soft sector, jet production by hard scattering, etc. so that users can employ a module based on their favorite physical description for each. 
For the in-medium parton shower evolution, the most distinctive feature of the \textsc{jetscape} framework is its support for multistage descriptions that, by stitching multiple models together, cover a broader range of scales. Depending on the virtuality or energy of a parton, each model becomes active to handle the parton shower evolution interactions with the medium. 

Recently, we systematically studied the energy loss of large-transverse momentum particles, jets, and charmed particles using a multistage model, combining two modules, \textsc{matter} for high-virtual parton shower and \textsc{lbt} for low virtuality, developed within the \textsc{jetscape} framework in Refs.~\cite{JETSCAPE:2022jer,JETSCAPE:2022hcb,JETSCAPE:2023ikg}. 
Our simulations indicate that the single high-$p_T$ particle spectra are dominated by the large virtuality phase simulated by the \textsc{matter} module. 
On the other hand, to describe the suppression of reconstructed jets and $D$ mesons, we found that the energy loss of soft daughter partons and heavy quarks is governed by the low-virtuality scattering dominated phase simulated by the \textsc{lbt} module. 

One further important insight from our prior work is that the reduction of the interaction with the medium at high virtuality due to modified coherence effects plays a crucial role in explaining the weak suppression of single charged particles with $p_{T}\gtrapprox 10~\mathrm{GeV}$. 
These modified coherence effects occur because the partons probing the medium have a small transverse size when the virtuality is large. 
A section of QGP resolved at such a shorter distance scale appears more dilute, resulting in fewer interactions~\cite{Kumar:2019uvu}.\footnote{In several other models, e.g., those in Refs.~\cite{Caucal:2019uvr,Ke:2020clc}, effects similar to modified coherence are implicitly introduced when jet particles tend to have large virtualities. 
They sharply turn off the medium effects in the early stages according to virtuality or other quantities. }

Modified coherence effects implemented in \textsc{matter} drastically improve the description of the transverse momentum dependence of the nuclear modification factor for inclusive single-charged particles, even at the qualitative functional behavior level. 
In contrast, for reconstructed jets at the currently available collision energies, modified coherence effects are not visible in the transverse momentum dependence of the nuclear modification factor, which only necessitated a readjustment of the overall medium coupling parameter $\alpha^{\mathrm{fix}}_{s}$. 
Thus, it is essential to search for the role of modified coherence effects in the evolution of jet showering patterns by examining further inner jet structure modification. 

In this paper, we systematically analyze the observables characterizing the internal structure of jets using the results of the exact same numerical simulations with \textsc{matter}+\textsc{lbt} that were used to study the nuclear modification factors for reconstructed jets and high $p_{T}$ single-charged particles in Ref.~\cite{JETSCAPE:2022jer}. The goal is to explore the details of the interaction strength at each scale on the internal structure of the jet. 
In particular, we examine the groomed jet observables, which display the effect of jet-medium interactions at the early high-virtuality stage, and the jet fragmentation function, which shows the medium effect on partons throughout a wide range of scales. 
In this work, we do not retune any parameters and employ those obtained in our previous work~\cite{JETSCAPE:2022jer}.

The paper is organized as follows. 
In Sec.~\ref{Section:Model}, salient characteristics of the underlying model are presented. 
In Subsec.~\ref{Subsection:ModelOverview}, an overview of the framework and setup is outlined. 
Subsection~\ref{Subsection:CoherenceEffects} is devoted to formulating modified coherence effects. 
This is followed by an investigation of the medium modification of jet substructure observables, focusing on modified coherence effects, by presenting results from our model calculations in Sec.~\ref{Section:Results}. Here, we also make predictions for the upcoming measurements of the jet substructure observables at RHIC. 
A summary of our results and concluding remarks are presented in Sec.~\ref{Section:Summary}. 
The {\hyperlink{Appen}{\textcolor{black}{Appendix}}} is dedicated to the presentation of our predictions of jet $R_{AA}$ at the top RHIC energy for benchmarking purposes.

\section{Model}
\label{Section:Model}

\textsc{jetscape} is a general-purpose event generator framework where different \emph{sub} event generators can be included in a modular fashion, producing an extensive end-to-end simulation of a heavy-ion collision. In this paper, we use the results of simulations that were generated in Ref.~\cite{JETSCAPE:2022jer} to calculate all jet substructure observables. This is not just for convenience but rather to demonstrate how the exact same simulations can simultaneously describe both the jet and leading hadron suppression, as well as several jet substructure observables. 

To that end, only a very brief overview of the components of the simulation will be provided in this section. The reader may refer to Ref.~\cite{JETSCAPE:2022jer} for specific details of the physics included in a \textsc{matter}+\textsc{lbt} simulation within the \textsc{jetscape} framework. Computational aspects of the \textsc{jetscape} framework are described in great detail in Ref.~\cite{Putschke:2019yrg}, while the basic physics of multistage simulators is described in Ref.~\cite{JETSCAPE:2017eso}.

\subsection{Overview}
\label{Subsection:ModelOverview}

To explore the medium modification of jet substructure, we perform simulations of jet events in high-energy nucleus-nucleus collisions utilizing the full framework of \textsc{jetscape} in two separate steps. 
First, we calculate the event-by-event space-time profiles of the QGP medium in nucleus+nucleus ($A+A$) collisions for the estimation of the local medium effect on parton shower evolution. 
For this part, we perform simulations of $(2+1)$-dimensional [$(2+1)$-D] free-streaming pre-equilibrium evolution~\cite{Liu:2015nwa} and subsequent viscous hydrodynamic evolution by the $(2+1)$-D \textsc{vishnu} code package~\cite{Shen:2014vra} with the initial condition generated by \textsc{trento}  \cite{Moreland:2014oya}. 
Here the maximum {\it a posteriori} (MAP) parameters obtained by Bayesian calibration in Ref.~\cite{Bernhard:2019bmu} are used for the Large Hadron Collider (LHC) energy calculations, while hand-tuned parameters were used for top RHIC energy. 

In the second step, the binary collision distribution from the same \textsc{trento}  initial condition as for the medium is used to sample the transverse position of a hard scattering. 
The hard scattering is produced by \pythia\ 8~\cite{Sjostrand:2019zhc} with initial-state radiation (ISR) and multiparton interaction (MPI) turned on, and final-state radiation (FSR) turned off. 
The produced partons in the hard scattering then undergo the multistage in-medium parton shower evolution within the \textsc{jetscape} framework. 
In this study, we use a combination of \textsc{matter} and \textsc{lbt} modules as described in Ref.~\cite{JETSCAPE:2022jer}.

The partons produced by hard scattering are first passed to the \textsc{matter} module, which simulates virtuality-ordered splitting of high-energy partons incorporating medium effects~\cite{Majumder:2013re,Cao:2017qpx}. 
This description by \textsc{matter} is valid for partons with virtuality sufficiently larger than the accumulated transverse momentum and virtuality generated by scattering from the medium. 

Partons whose virtuality is reduced by showering in \textsc{matter} are then transferred to \textsc{lbt} at a transition scale. 
In \textsc{lbt}, the kinetic theory for on-shell partons with elastic and inelastic scatterings with medium constituents is applied~\cite{Wang:2013cia,He:2015pra,Cao:2016gvr}. 
The parton splittings under this description are entirely scattering-driven. 
In the multistage approach of the \textsc{jetscape} framework, virtuality-dependent switching between modules is done bidirectionally on a per-parton basis using a switching parameter $Q^2_{\mathrm{sw}}$. 
If the virtuality of the parton $Q^2 = p^\mu p_\mu - m^2$ falls below $Q^2_{\mathrm{sw}}$, it is then sent from \textsc{matter} to \textsc{lbt}. 
Conversely, the parton is returned to \textsc{matter} if its virtuality exceeds $Q^2_{\mathrm{sw}}$ again, or it goes out of the dense medium. The transition from medium-like back to vacuum-like emission takes place at a boundary with a temperature $T_{c} = 0.16$~GeV. 
In this study, $Q^2_{\mathrm{sw}}$ is set to $4~{\mathrm{GeV}}^2$. 

\pagelabel{rererevision-hadronization-model-limit}
After all the partons are outside the QGP medium and have virtuality smaller than the cutoff scale $Q_{\mathrm{min}}^2=1\,{\mathrm{GeV}}^2$, they are hadronized via the Colorless Hadronization module, in which the Lund string model of \pythia\ 8 is utilized. 
In this study, we do not yet consider medium effects on the hadronization process, such as those modeled by the Hybrid Hadronization approach~\cite{Han:2016uhh,Fries:2019vws,JETSCAPE:2023ewn:Manual}. While these effects are expected to be particularly significant in observables focused on the medium to low momentum regions, a more systematic investigation of this is left for future work. 

In both \textsc{matter} and \textsc{lbt} modules, the medium response effect is taken into account via recoil partons~\cite{Li:2010ts,Zapp:2012ak,Zapp:2013vla,Cao:2017hhk,Luo:2018pto,Park:2018acg,Tachibana:2020mtb}. 
In the \emph{recoil} prescription, the energy-momentum transfer is described by scatterings between jet partons and medium partons. 
For each scattering, a parton is sampled from the thermal medium. 
Then, the scattered sampled parton is assumed to be on-shell, and passed to \textsc{lbt} for its in-medium evolution, assuming weak coupling with the medium. 
These \emph{recoil} partons and further accompanying daughter partons are collectively hadronized with the other jet shower partons. 
On the other hand, a deficit of energy and momentum in the medium is left for each recoil process, where a parton emanating from the medium is included, postscattering, as a part of the jet. 
We treat this deficit as a freestreaming particle, referred to as a \emph{hole} parton, and track it. 
The hole partons are hadronized separately from other jet partons, and their energy and momentum within each positive particle jet cone are subtracted in the jet clustering routine to ensure energy-momentum conservation~\cite{Luo:2023nsi}.

\pagelabel{rererevision-perturbation-model-limit}
In this study, the description of parton shower evolution and its interaction with the medium is limited to a perturbative framework. While interactions at nonperturbative scales would require the inclusion of hydrodynamic responses, as will be explained shortly, the model does not incorporate these effects. Therefore, to ensure consistency and facilitate systematic study, the description of parton shower evolution is restricted to the perturbative regime. 

\pagelabel{rererevision-no-hydro-res-model-limit}
In the later stages, where the energy of a jet shower parton reaches a comparable scale to the ambient temperature, the mean-free path is no longer large enough to apply the kinetic theory-based approach with the recoil prescription. In principle, such soft components of jets are supposed to be thermalized and evolve hydrodynamically as part of the bulk medium fluid~\cite{Casalderrey-Solana:2004fdk,Stoecker:2004qu,Tachibana:2019hrn,Cao:2020wlm,Schlichting:2020lef,Luo:2021iay,Mehtar-Tani:2022zwf:Manual}. As in Refs.~\cite{Chaudhuri:2005vc,Renk:2005si,Satarov:2005mv,Neufeld:2008fi,Noronha:2008un,Qin:2009uh,Betz:2010qh,Neufeld:2011yh,Schulc:2014jma,Tachibana:2014lja,JETSCAPE:2020uew}, implementation of models based on such a description is proposed, and there are some studies of the hydrodynamic medium response to jets using it~\cite{Tachibana:2017syd,Okai:2017ofp,Chen:2017zte,Chang:2019sae,Tachibana:2020mtb,Casalderrey-Solana:2020rsj,Yang:2021qtl,Yang:2022nei,Pablos:2022piv,Yang:2022yfr}. 
However, with such an implementation of the hydrodynamic medium response, the computational cost for a systematic and exhaustive study covering various configurations, as presented in this paper, is enormously expensive. Thus, in this paper, we mainly discuss the structure of the hard part of the jet, where the contributions of such very soft components are relatively small. A further comprehensive investigation with more detailed modeling of the medium response in jet modification is left for future work.

To investigate the modification of jet substructures by medium effects in $A+A$ collisions, the calculations of the same observables for $p$+$p$ collisions are necessary as references. 
For such calculations, the parton shower evolution modules are replaced entirely by \textsc{matter} with no in-medium scattering. 
This setup for $p$+$p$ collisions of \textsc{jetscape}, referred to as the \textsc{jetscape} PP19 tune, is equivalent to the limit of no medium effect in the event and is detailed in Ref.~\cite{JETSCAPE:2019udz}.

\subsection{Modified coherence effects at high virtuality}
\label{Subsection:CoherenceEffects}

In this study, we focus on \emph{modified coherence} effects~\cite{Kumar:2019uvu}.  While these are similar to what is typically referred to as color coherence effects in the literature~\cite{Mehtar-Tani:2010ebp,Mehtar-Tani:2011hma,Casalderrey-Solana:2011ule,Caucal:2018dla}, there are some differences. 

\pagelabel{rerevision-comment1-coherence}
In typical color coherence calculations, soft gluons from the medium cannot resolve the hard antenna formed in VLE and so cannot affect any energy loss until partons in the antenna have separated to a resolvable distance. Along with this, soft, large angle radiations from this small antenna are suppressed due to destructive interference~\cite{Dokshitzer:1987nm}. As a result, the color antenna develops in the medium as it does in the vacuum until colored partons are sufficiently separated even after the formation time. 

By contrast, modified coherence should be taken into account during the formation time of the highly virtual parton that is about to undergo VLE. Short distance color fluctuations in the medium can resolve this virtual splitting within the formation time. Indeed, according to the higher twist formalism~\cite{Wang:2001ifa,Majumder:2009ge,Qin:2009gw}, such interactions between virtual emitted gluons and the medium constitute the primary contribution to medium modifications on VLE. However, as the virtuality of the partons increases, the size of the virtual splitting becomes smaller and harder to resolve, leading to such interactions progressively diminishing. We refer to this process, where the medium effects on virtuality-driven VLE are suppressed depending on the size of the jet parton's virtuality, as modified coherence. 

\pagelabel{rererevision-coherence-comparison}
It should be noted that typical color coherence and modified coherence are not in a relationship where one completely encompasses the other, while they might consequently lead to a similar effect under most circumstances. 
In our \textsc{jetscape} simulations presented in this paper, whether the setup incorporates modified coherence or not, typical color coherence is not included. 

Modified coherence can be incorporated as in Ref.~\cite{Kumar:2019uvu}. In that reference, it was demonstrated that a hard parton with large virtuality resolves the very short-distance structure of the medium via the exchange of a gluon whose momentum is much larger than the medium temperature. 
These modified coherence effects are formulated with the continuous evolution of the medium-resolution scale and give a gradual reduction of jet parton-medium interaction as a function of the virtuality. 

For jet quenching calculations, modified coherence effects can be effectively implemented by introducing a modulation factor $f(Q^2)$, which diminishes as a function of the parent parton's virtuality $Q^2$, in the medium-modified splitting function: 
\begin{multline}
\Tilde{P}_a(y,Q^2)=P^{\mathrm{vac}}_a(y)\\
\times
\left\{ 1 + 
\int\limits^{\tau_{\mathrm{form}}^{+}}_{0} 
d\xi^{+}
\hat{q}^a_{\mathrm{HTL}}
\frac{c^{a}_{\hat{q}}  f(Q^2) \left[ 2 - 2\cos \left( \frac{\xi^{+} }{\tau_{\mathrm{form}}^{+}} \right) \right]}{y(1-y) Q^{2} (1+\chi_a)^{2}} \right\}.
\label{eq:HT-splitting function} 
\end{multline}
In the equation above, $P^{\mathrm{vac}}_a(y)$ is the Altarelli-Parisi vacuum splitting function~\cite{Altarelli:1977zs} for the parent parton species $a=(g,q,\bar{q})$ with the forward light-cone momentum fraction of the daughter parton $y$, 
$\chi_a=(\delta_{aq}+\delta_{a\bar{q}})y^2m_a^2/[y(1-y)Q^2 -y^2m_a^{2}]$ with $m_a$ being the parent parton mass, and $c^{a}_{\hat{q}}=\left[1-\frac{y}{2}\left(\delta_{a,q}+\delta_{a,\bar{q}}\right)\right] - \chi_a \left[1 - \left(1-\frac{y}{2}\right) \chi_a\right]$. 
The integration in Eq.~(\ref{eq:HT-splitting function}) is taken over light-cone time $\xi^+$ with the upper bound $\tau_{\mathrm{form}}^{+}=2p^{+}/Q^{2}$ being the formation time of the radiated parton, where $p^+=p^\mu \hat{n}_{\mu}/\sqrt{2}$ [with $\hat{n}_{\mu}=\left(1,\mathbf{p}/\lvert \mathbf{p}\rvert\right)$] is the forward light-cone momentum of the parent parton. 
The formulation of $\Tilde{P}_{a}(y,Q^2)$ in Eq.~(\ref{eq:HT-splitting function}) is obtained using soft collinear effective theory within the higher twist scheme~\cite{Abir:2014sxa,Abir:2015hta}. 

The parameterization of the virtuality-dependent modulation factor is given as~\cite{Amit2020PhDThesis,JETSCAPE:2022jer}
\begin{align}
f(Q^2) & = 
\begin{cases}
\frac{1+10\ln^{2}(Q^2_\mathrm{sw}) + 100\ln^{4}(Q^2_\mathrm{sw})}{1+10\ln^{2}(Q^2) + 100\ln^{4}(Q^2)} & 
\text{if } Q^2 > Q_{\rm sw}^2 \\ 
1 & \text{if } Q^2 \le Q_{\rm sw}^2 
\end{cases}. 
\label{eq:qhatSuppressionFactor}
\end{align}
\pagelabel{rerevision-comment1-modulation-factor}
As presented in Ref.~\cite{JETSCAPE:2022jer}, this function sharply decreases as the virtuality increases, e.g., $f(Q^2\!=\!(5~\mathrm{GeV})^{2})\!\approx\! 0.0359$, $f(Q^2\!=\!(10~\mathrm{GeV})^{2})\!\approx\! 0.00862$, and $f(Q^2\!=\!(50~\mathrm{GeV})^{2})\!\approx\! 0.00104$. 
When this explicit virtuality dependence is eliminated, the strength of the medium effect is controlled solely by the conventional transport coefficient for a low virtuality (near on shell) parton from the hard-thermal-loop (HTL) calculation~\cite{He:2015pra,Caron-Huot:2010qjx},
\begin{align}
\hat{q}^{a}_{\mathrm{HTL}} =C_{a}\frac{42 \zeta(3)}{\pi}  \alpha^{\mathrm{run}}_{s}(2p^0 T) \alpha^{\mathrm{run}}_{s} (m_{D}^2)T^{3} \ln\left[\! \frac{2p^0T}{m^2_D}\! \right]. 
\label{eq:HTL-qhat-formula-C-2}
\end{align}
Here, 
$C_{a}$ is the Casimir color factor for the hard parent parton, 
$\zeta(3)\approx 1.20205$ is Ap\'{e}ry's constant, 
$p^0$ is the energy of the hard parent parton, 
$T$ is the local temperature, 
and 
$m^2_{D}=\frac{4\pi\alpha^{\mathrm{run}}_s(m_{D}^2) T^2}{3} \left(N_{c}+\frac{N_{f}}{2}\right)$ is the Debye screening mass for a QCD plasma with $N_{c}= 3$ colors and $N_{f}= 3$ fermion flavors. 

The running coupling constant is 
\begin{align}
\alpha^{\mathrm{run}}_{s}(\mu^2)
&=
\begin{cases}
\frac{4\pi}{11-2N_{f}/3} \frac{1}{\ln\left(\mu^2/\Lambda^2\right)}
& \text{if } \mu^2 > \mu^2_0\\
\alpha^{\mathrm{fix}}_{s} & 
\text{if } \mu^2 \leq \mu^2_0 
\end{cases},
\label{eq:running_coupling}
\end{align}
with $\Lambda$ being chosen such that $\alpha^{\mathrm{run}}_{s}(\mu_0^2)=\alpha^{\mathrm{fix}}_{s}$ at $\mu_0^2=1~\mathrm{GeV}^2$. 
A similar implementation of the running coupling that appears in Eq.~\eqref{eq:HTL-qhat-formula-C-2} can be found, for example, in the case of 2-2 scattering in Ref~\cite{Peigne:2008nd}. 
For most temperature regions accessible at RHIC and LHC collision energies, $m^2_D \lessapprox \mu^2_{0}=1~\mathrm{GeV}^2$, and therefore $\alpha_s(m^2_D) = \alpha^{\mathrm{fix}}_s$. 
In this framework, $\alpha^{\mathrm{fix}}_{s}$ is the free parameter controlling the overall interaction strength and chosen to give the best fit to the experimental data of inclusive jet $R_{AA}$~\cite{JETSCAPE:2022jer}.

\pagelabel{rerevision-comment3-params}
In this paper, we compare results from
two different setups: with the virtuality-dependent modified coherence effects and without any coherence effects (referred to as Type-3 and Type-2 in Ref.~\cite{JETSCAPE:2022jer}, respectively). 
For the case with modified coherence, $\Tilde{P}_a(y,Q^2)$ in Eq.~(\ref{eq:HT-splitting function}), with the virtuality-dependent modulation factor from Eq.~(\ref{eq:qhatSuppressionFactor}), is employed in the high virtuality phase by \textsc{matter}, with $\alpha^{\mathrm{fix}}_{s}=0.3$.\footnote{This configuration for \textsc{matter}+\textsc{lbt} with modified coherence effects is referred to as JETSACPEv3.5 AA22 tune, and its results are provided as defaults for comparisons with experimental and other data.} 
In the setup without coherence effects, the modulation factor is fixed to unity [$f(Q^2)=1$] for any $Q^2$ to eliminate the explicit virtuality dependence. 
The best fit with leading hadron and jet data is obtained with an $\alpha^{\mathrm{fix}}_{s}=0.25$ for this case. 
We present results for jet substructure using events generated with the above parametrizations, for both setups. 

\section{Results}
\label{Section:Results}
In this section, we present the results 
for jet substructure observables in Pb$+$Pb collisions at $\sqrt{s_{\mathrm{NN}}}=5.02~\mathrm{TeV}$ based on the multistage (\textsc{matter}+\textsc{lbt}) jet quenching model described in the previous section.
A complementary study of the nuclear modification factor $R_{AA}$ for reconstructed jets and charged particles using the same model has been presented in Ref.~\cite{JETSCAPE:2022jer}. Moreover, this same formalism has been applied to study the  heavy-flavor observables and has been presented in Ref.~\cite{JETSCAPE:2022hcb,JETSCAPE:2023ikg}.

To show the capability of the \textsc{jetscape} framework, we also provide predictions of the groomed jet observables, fragmentation function, and jet cone size dependence of inclusive jets and charged jets for the upcoming jet measurements at RHIC.
Throughout this work, the jet reconstruction and Soft Drop grooming are performed using the \textsc{fastjet} package~\cite{Cacciari:2005hq, Cacciari:2011ma} with \textsc{fastjet-contrib}~\cite{fjcontrib_code}. 

\subsection{Groomed jet observables}
In this section, we present the observables obtained via Soft Drop grooming algorithm~\cite{Larkoski:2014wba,Dasgupta:2013ihk,Larkoski:2015lea}. 
The Soft Drop procedure removes the contributions from soft wide-angle radiation and enables access to the hard parton splittings during the jet evolution. 
In this algorithm, first, jets are constructed by a standard jet finding algorithm such as the anti-$k_{t}$ algorithm~\cite{Cacciari:2008gp} with a definite jet cone size $R$. Then, the constituents of an anti-$k_{t}$ jet are again reclustered by the Cambridge-Aachen (C/A) algorithm~\cite{Dokshitzer:1997in,Wobisch:1998wt} to form a pairwise clustering tree. 
The next step is to trace back the C/A tree. Here, one declusters the C/A jet by undoing the last step of the C/A clustering and selecting the resulting two prongs. The two prongs are checked to see if they satisfy the Soft Drop condition, given as:
\begin{align}
\label{eq:soft_drop}
\frac{\min\left(p_{T,1},
p_{T,2}\right)}{p_{T,1}+
p_{T,2}}>z_{\mathrm{cut}}\left(\frac{\Delta R_{12}}{R}\right)^{\beta}, 
\end{align}
where $p_{T,1}$ and $p_{T,2}$ are the transverse momenta of the prongs, 
$\Delta R_{12} = [(\eta_{1}-\eta_{2})^{2}
+(\phi_{1}-\phi_{2})^{2}]^{1/2}$ is the radial distance between the prongs in the rapidity-azimuthal angle plane, $z_{\mathrm{cut}}$ and $\beta$ are parameters controlling the grooming procedure. 
If the condition is failed, the prong with the larger $p_{T}$ of the pair is further declustered into a pair of prongs. 
This process is repeated until one finds a pair of prongs satisfying the Soft Drop condition. The resulting pair of prongs are used to compute the groomed jet observables.
It is worth noting that there may exist cases in which no prong pair passing the soft-drop condition is eventually found even if the C/A tree is traversed back to the end; such cases are referred to as ``Soft Drop fail.''

\begin{figure*}[htb!]
\centering
\includegraphics[width=0.98\textwidth]{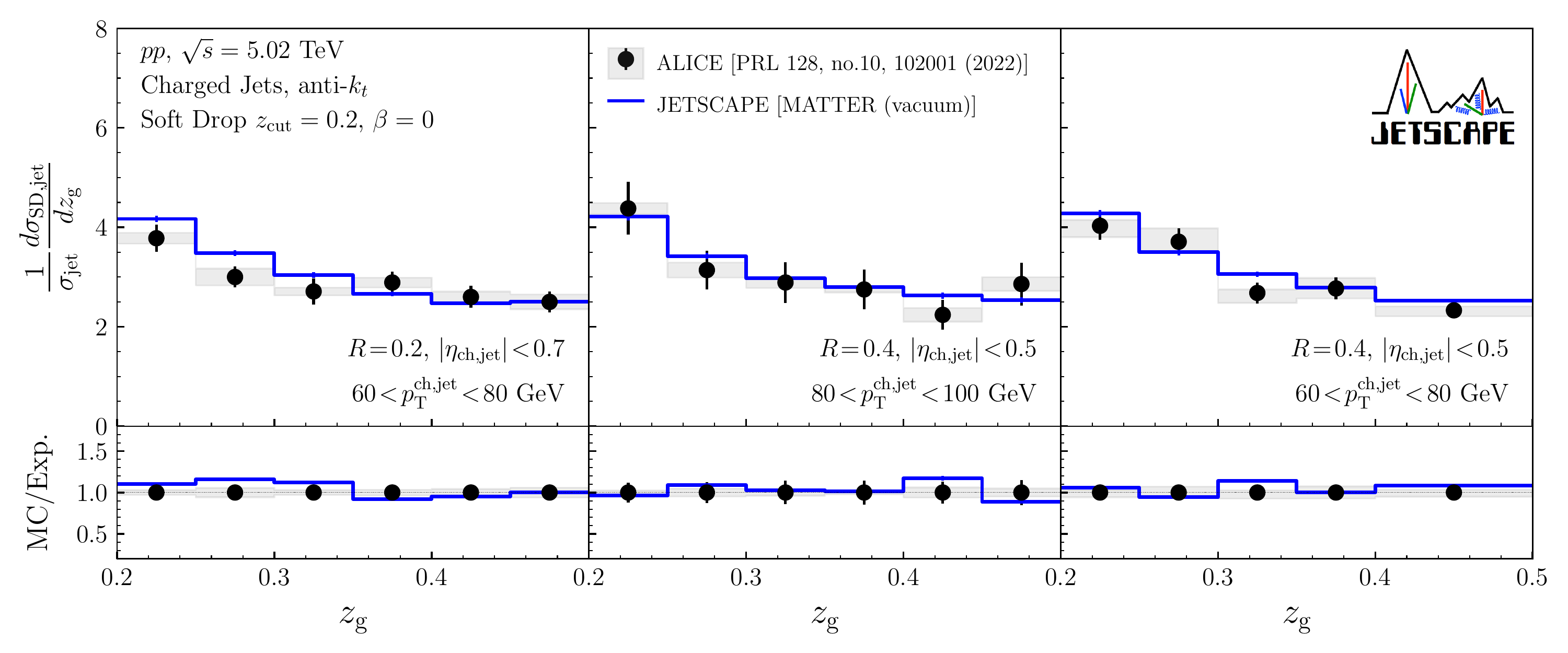}
\caption{(Color online) Distributions of jet splitting momentum fraction $z_{g}$ for charged jets 
in $p$+$p$ collisions at $\sqrt{s}=5.02$~TeV and the ratios 
for different jet cone size $R$, and $p^{\mathrm{ch,jet}}_{T}$ range. 
The Soft Drop parameters are $z_{\mathrm{cut}}=0.2$ and $\beta = 0$. 
The solid lines and circles with statistical error bars show the results from \textsc{jetscape} and the experimental data from ALICE Collaboration~\cite{ALargeIonColliderExperiment:2021mqf}, respectively. 
The bands indicate the systematic uncertainties of the experimental data. 
}
\label{fig:alice_zg_pp}
\end{figure*}

\subsubsection{Jet splitting momentum fraction}
Here, we study the medium modification of the jet splitting momentum fraction $z_{g}$, which is defined as the left-hand side of Eq.~(\ref{eq:soft_drop}) in the case with the prong pair passing the Soft Drop condition.

\begin{figure*}[htbp]
\centering
\includegraphics[width=0.98\textwidth]{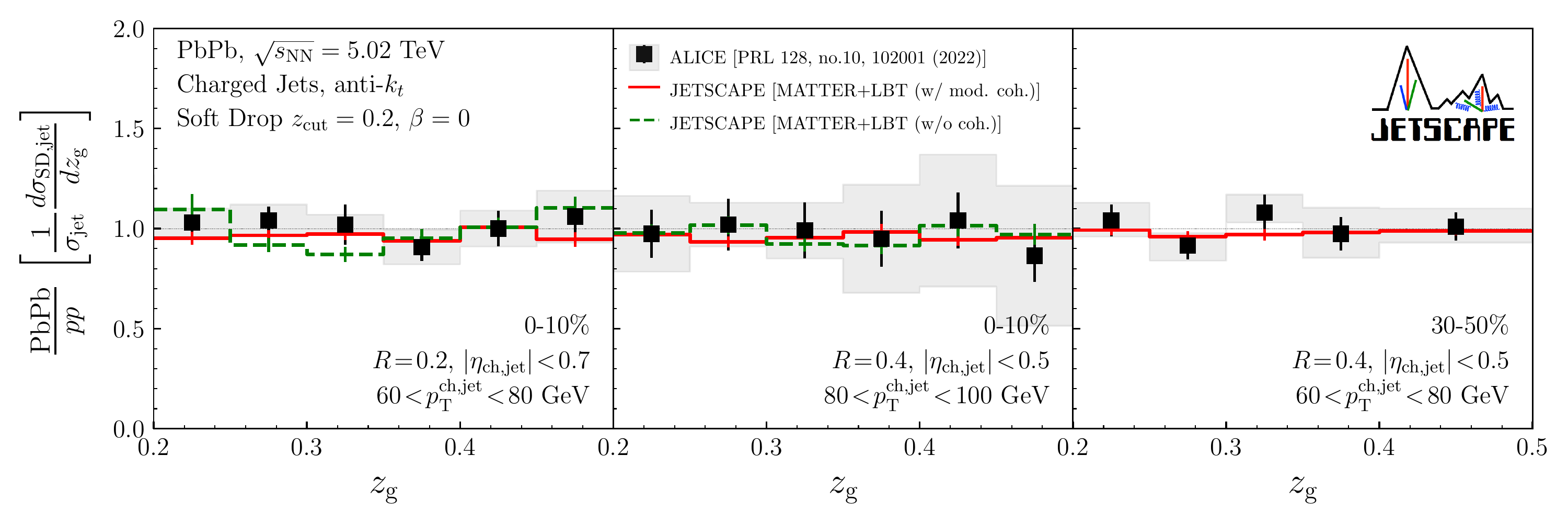}
\caption{(Color online) 
Ratios of $z_{g}$ distributions for charged jets 
between Pb$+$Pb and $p$+$p$ collisions at $\sqrt{s_{\mathrm{NN}}}=5.02$~TeV
for different centrality, jet cone size $R$, and $p^{\mathrm{ch,jet}}_{T}$ range. 
The Soft Drop parameters are $z_{\mathrm{cut}}=0.2$ and $\beta = 0$. 
The solid and dashed lines with statistical error bars show the results from \textsc{matter}+\textsc{lbt} of \textsc{jetscape} with modified coherence effects and without any coherence effects, respectively. 
For comparison, the experimental data from the ALICE Collaboration~\cite{ALargeIonColliderExperiment:2021mqf} 
are shown by squares with statistical errors (bars) and systematic uncertainties (bands). }
\label{fig:alice_zg_pbpb}
\end{figure*}

Figure~\ref{fig:alice_zg_pp} shows $z_{g}$ distributions for charged jets in $p$+$p$ collisions at $\sqrt{s}=5.02$~TeV defined as 
\begin{align}
\label{eq:zg_distribution_alice}
\frac{1}{\sigma_{\mathrm{jet}}}\frac{d\sigma_\mathrm{SD,jet}}{dz_{g}}&=\frac{1}{N_{\mathrm{jet}}}\frac{dN_\mathrm{SD,jet}}{dz_{g}}, 
\end{align}
where $N_{\mathrm{jet}}$ is the number of inclusive jets, 
$N_\mathrm{SD,jet}$ is the number of jets passing the Soft Drop condition 
and $\sigma_{\mathrm{jet}}$, $\sigma_\mathrm{SD,jet}$ are the corresponding cross sections. 
The Soft Drop parameters are set as $z_{\mathrm{cut}} = 0.2$ and $\beta = 0$. 
The results from the \textsc{jetscape} PP19 tune for different $p^{\mathrm{ch,jet}}_{T}$ ranges and jet cone sizes 
are compared with the experimental data from ALICE. 
Some small discrepancies can be seen, but they are mostly compatible within uncertainty. 

\begin{figure*}[htbp]
\centering
\includegraphics[width=0.75\textwidth]{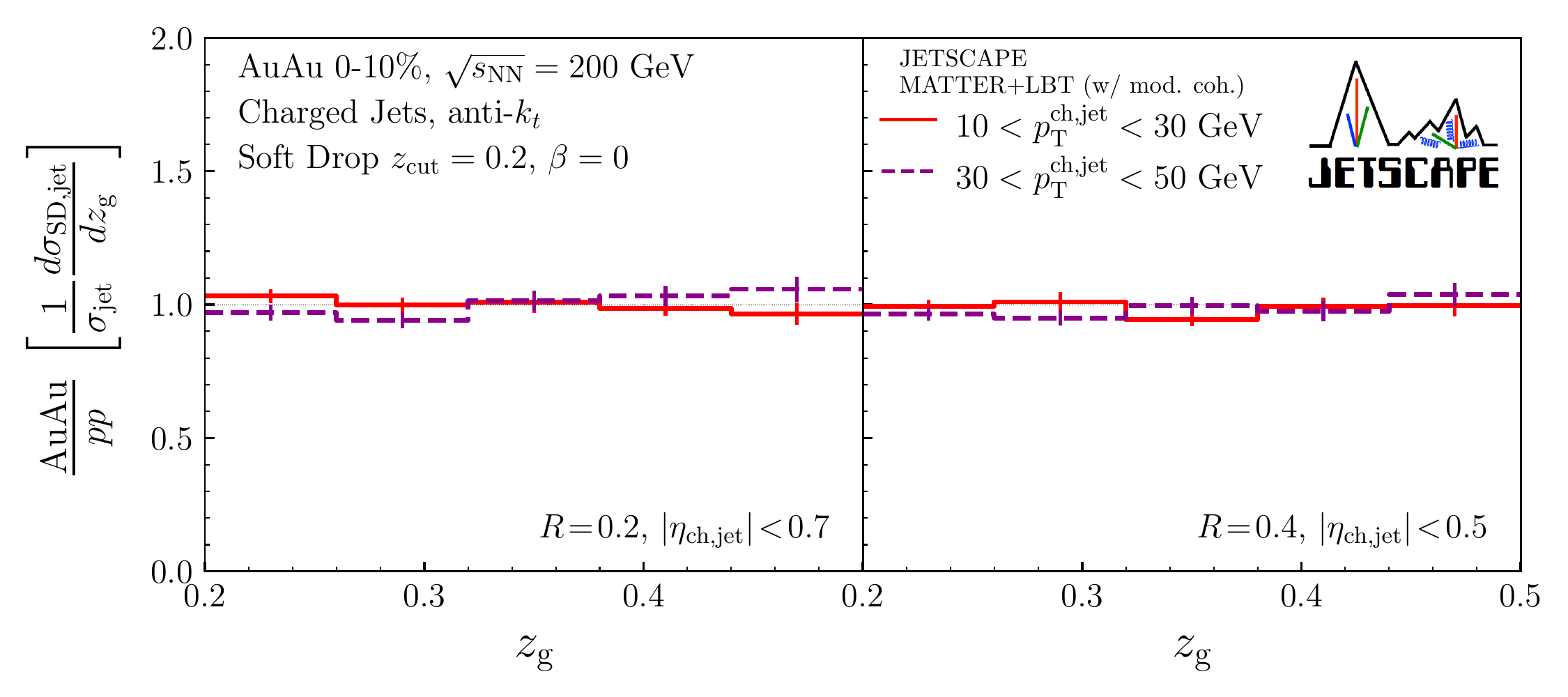}
\caption{(Color online) 
Ratios of $z_{g}$ distributions for charged jets with $R=0.2$ and $|\eta_{\mathrm{ch,jet}}|<0.7$ (left), and $R=0.4$ $|\eta_{\mathrm{ch,jet}}|<0.5$ (right) between $0\%$--$10\%$ Au$+$Au and $p$+$p$ collisions at $\sqrt{s_{\mathrm{NN}}}=200$~GeV from \textsc{matter}+\textsc{lbt} of \textsc{jetscape} with modified coherence effects. 
The Soft Drop parameters are $z_{\mathrm{cut}}=0.2$ and $\beta = 0$. 
The solid and dashed lines with statistical error bars show the results for $10<p^{\mathrm{ch,jet}}_{T}<30$~GeV and $30<p^{\mathrm{ch,jet}}_{T}<50$~GeV, respectively. 
}
\label{fig:rhic_zg}
\end{figure*}

In Fig.~\ref{fig:alice_zg_pbpb}, the modification of the $z_{g}$ distribution for charged jets is presented as the ratio of the distribution in Pb$+$Pb to $p+p$ collisions at $\sqrt{s_{\mathrm{NN}}}=5.02$~TeV. 
Both results, with and without consideration of modified coherence effects, do not exhibit significant modification and are consistent with the experimental data. 
This indicates that the medium effects on the functional form for the momentum fraction $y$ of the splitting function are small in hard partonic splittings. To be clear, the entire ensemble of jets in Pb$+$Pb that are included in this analysis is indeed modified by the medium. 
\pagelabel{rerevision-comment2-zg-mod}
However, most modifications occur at softer momenta in jets, leaving the hard splittings unaffected.  

Next, for upcoming measurements at RHIC, we present the prediction of the modification of the $z_{g}$ distribution for charged jets in $0\%$--$10\%$ Au$+$Au collisions at $\sqrt{s_{\mathrm{NN}}}=200$~GeV from \textsc{matter}+\textsc{lbt} with modified coherence effects in Fig.~\ref{fig:rhic_zg}. The trend is similar to the results observed at the LHC collision energy and does not show any significant nuclear effects for the kinematic configurations considered. 

\subsubsection{Jet splitting radius}
\label{Section:SplittingRadius}
\begin{figure*}[htb!]
\centering
\includegraphics[width=0.98\textwidth]{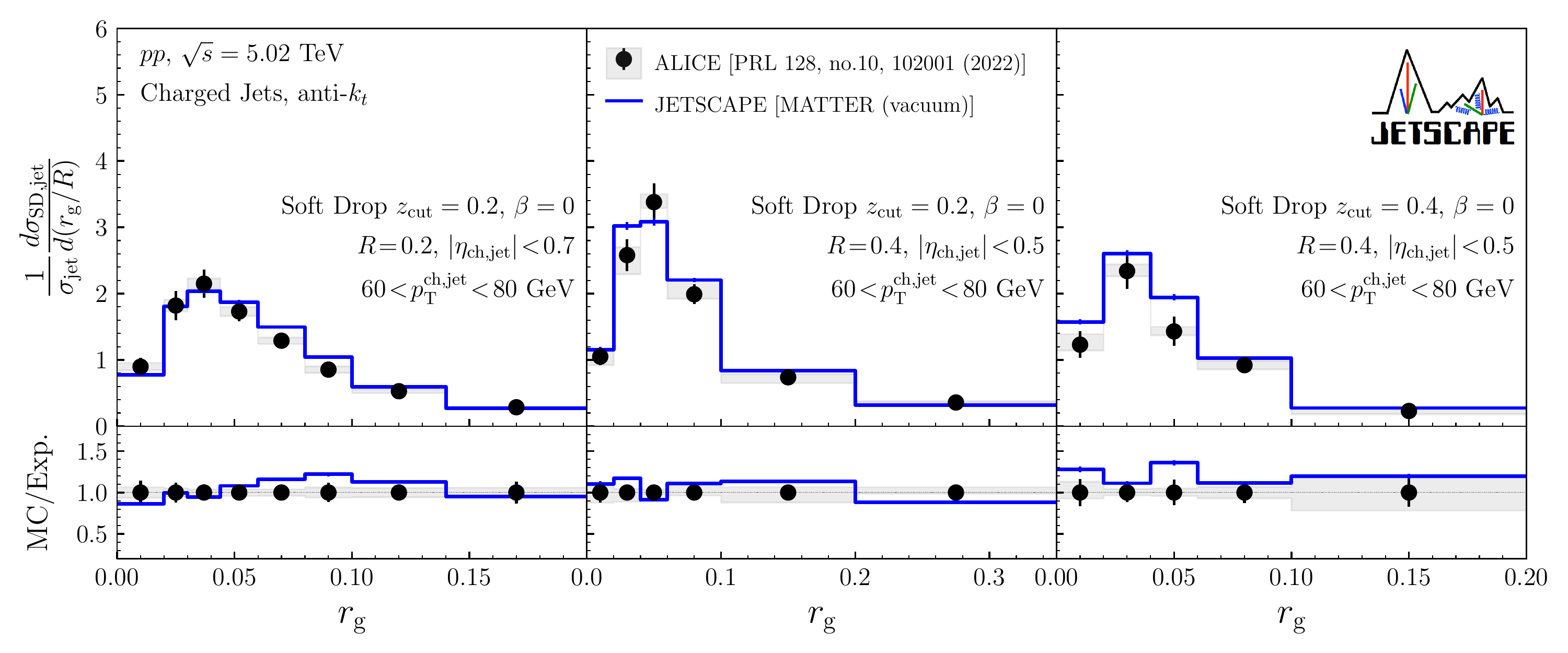}
\caption{(Color online) Distributions of jet splitting radius $r_{g}$ for charged jets in $p$+$p$ collisions at $\sqrt{s}=5.02$~TeV and the ratios 
for different jet cone size $R$, and $p^{\mathrm{ch,jet}}_{T}$ range. 
The Soft Drop parameters are $z_{\mathrm{cut}}=0.2$ and $\beta = 0$. 
The solid lines and circles with statistical error bars show the results from \textsc{jetscape} and the experimental data from the ALICE Collaboration~\cite{ALargeIonColliderExperiment:2021mqf}, respectively. 
The bands indicate the systematic uncertainties of the experimental data. }
\label{fig:alice_rg_pp}
\end{figure*}

Next, we study the medium modification of jet splitting radius $r_{g}$, defined as the radial distance $\Delta R_{12}$ of the prong pair passing the Soft Drop condition. 
In Fig.~\ref{fig:alice_rg_pp}, $r_{g}$ distributions defined as 
\begin{align}	
\frac{1}{\sigma_{\mathrm{jet}}}\frac{d\sigma_\mathrm{SD,jet}}{d\left(r_{g}/R\right)}&=\frac{1}{N_{\mathrm{jet}}}\frac{dN_\mathrm{SD,jet}}{d\left(r_{g}/R\right)}, 
\end{align}
are shown for charged jets in $p$+$p$ collisions at $\sqrt{s}=5.02$~TeV. 
The results from the \textsc{jetscape} PP19 tune show good agreement with the ALICE data, particularly for the cases with $z_{\mathrm{cut}}=0.2$.

\begin{figure*}[!htb]
\centering
\includegraphics[width=0.98\textwidth]{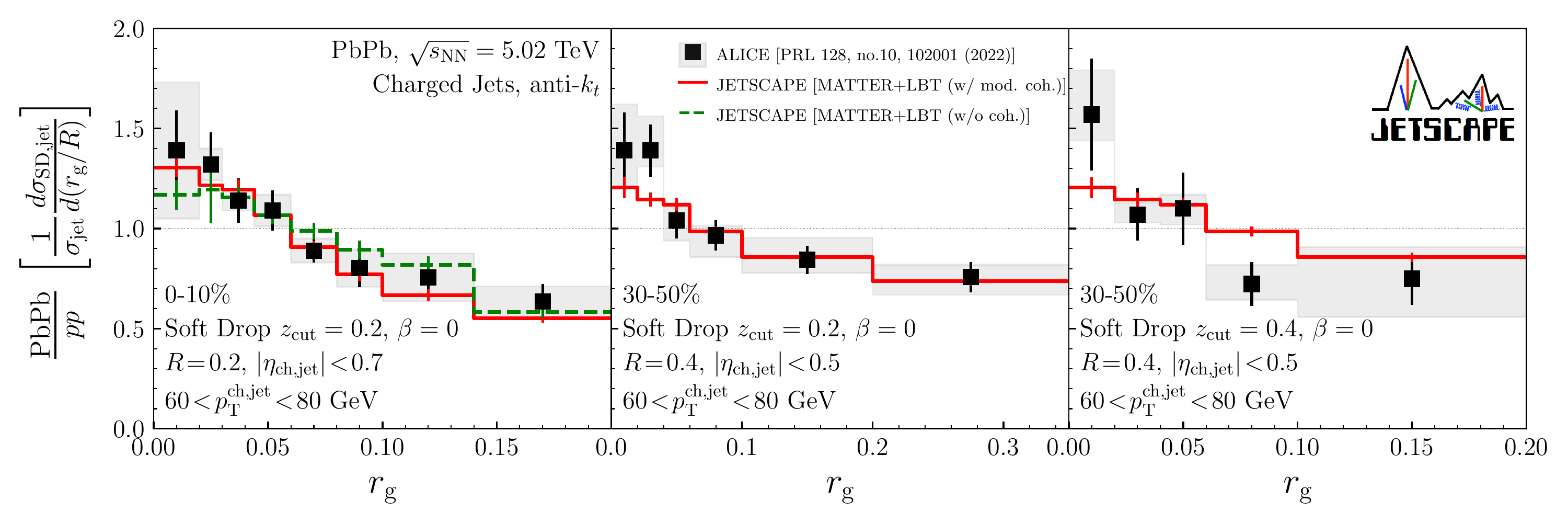}
\caption{(Color online) 
Ratios of $r_{g}$ distributions for charged jets 
between Pb$+$Pb and $p$+$p$ collisions at $\sqrt{s_{\mathrm{NN}}}=5.02$~TeV
for different centrality, jet cone size $R$, soft drop parameter $z_{\mathrm{cut}}$, and $p^{\mathrm{ch,jet}}_{T}$ range. 
The solid and dashed lines with statistical error bars show the results from \textsc{matter}+\textsc{lbt} of \textsc{jetscape} with modified coherence effects and without any coherence effects, respectively. 
For comparison, the experimental data from the ALICE Collaboration~\cite{ALargeIonColliderExperiment:2021mqf} 
are shown by squares with statistical errors (bars) and systematic uncertainties (bands). }
\label{fig:alice_rg_pbpb}
\end{figure*}

\pagelabel{rererevision-zcut-04}
Figure~\ref{fig:alice_rg_pbpb} shows the modification of $r_{g}$ distribution for charged jets in Pb$+$Pb collisions at $\sqrt{s_{\mathrm{NN}}}=5.02$~TeV. 
Our full results with modified coherence effects capture the trend observed in experimental data: Enhancement at small $r_g$ and suppression at large $r_g$. 
In particular, the agreements within uncertainties can be seen for the case with $z_{\mathrm{cut}} = 0.2$. 
For the case with $z_{\mathrm{cut}} = 0.4$, it is acknowledged that the description by the results with modified coherence may be somewhat inadequate, as the modification pattern is likely underestimated. However, it is also important to note that the condition with $z_{\mathrm{cut}} = 0.4$ involves a very small phase-space region, making the analysis more stringent, and the experimental data currently have larger error bars. With experimental results with smaller uncertainty in the future, more quantitative discussions can be conducted, allowing for a more detailed investigation. 

For the $0\%$--$10\%$ most central bin, the result without coherence effects is shown for comparison. 
It gives a slightly smaller slope, but no conclusion can be drawn within the current uncertainties. 
Combined with the results for the $z_g$ distribution, we obtain the clear conclusion that these jets passing the Soft Drop condition are indeed modified, but predominantly in their softer components rather than in the hard partonic splittings. 
For jets originally having a larger hard-splitting angle, the soft component diffusing due to the medium effect is more likely to leave the jet cone, resulting in more considerable energy loss. 
Thus, jets with larger hard splitting angles are less likely to be triggered, and the narrowing is observed as the yield ratio of jets with smaller splitting angles increases. 

\begin{figure*}[htb!]
\centering
\includegraphics[width=0.98\textwidth]{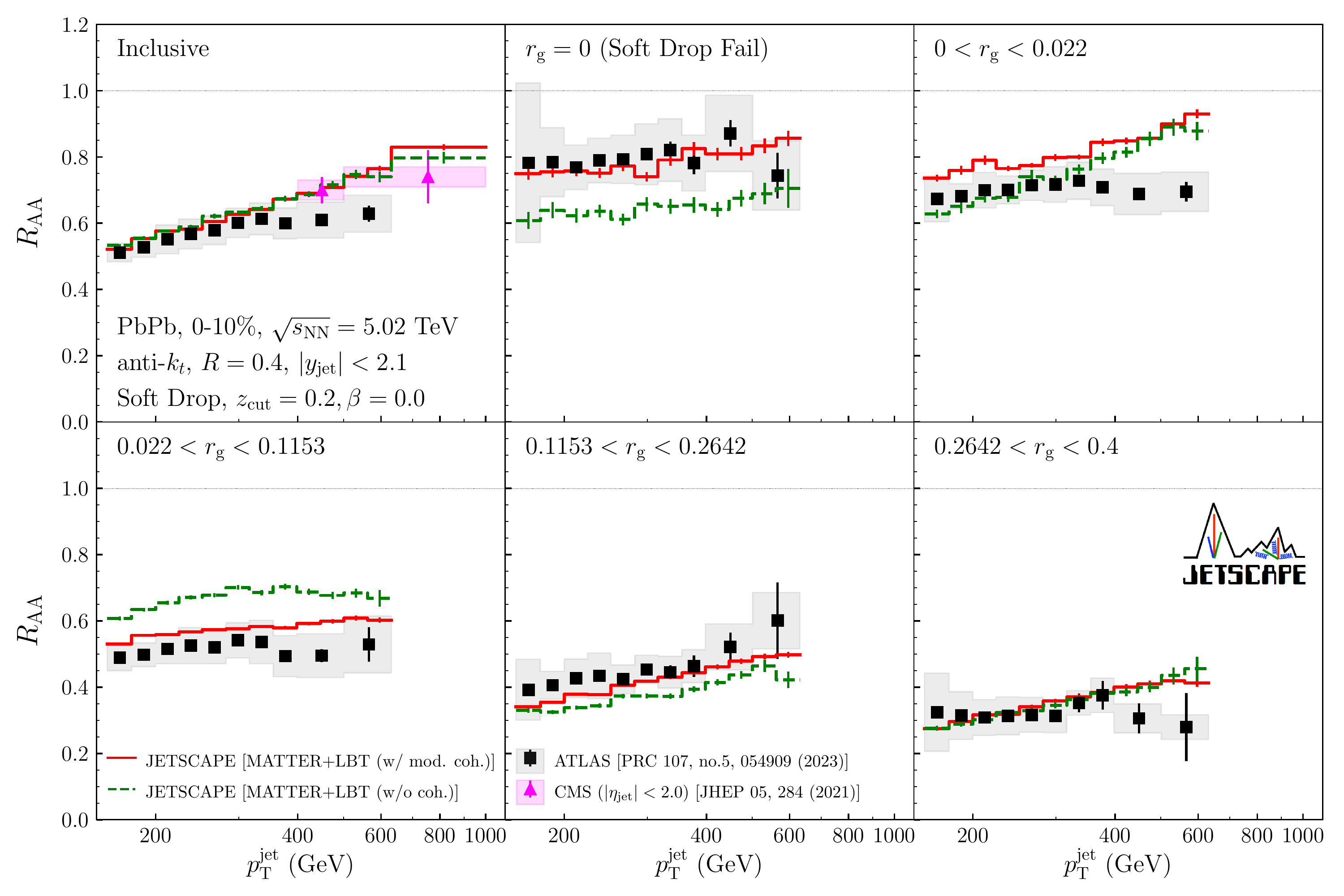}
\caption{(Color online) Nuclear modification factor $R_{AA}$ as a function of $p_{T}^{\mathrm{jet}}$ for inclusive jets,
jets failing the Soft Drop condition ($r_{g}=0$), and groomed jets with different splitting radius $r_{g}$ in $0\%$--$10\%$ Pb$+$Pb collisions at $\sqrt{s_{\mathrm{NN}}}=5.02$~TeV. 
Jets are reconstructed with $R=0.4$ at midrapidity $|y_{\mathrm{jet}}|<2.1$. 
The Soft Drop parameters are $z_{\mathrm{cut}}=0.2$ and $\beta = 0$. 
The solid and dashed lines with statistical error bars show the results from \textsc{matter}+\textsc{lbt} of \textsc{jetscape} with modified coherence effects and without any coherence effects, respectively. 
The results are compared with ATLAS data~\cite{ATLAS:2022vii} (squares) 
and CMS data for  $\left\vert\eta_{\mathrm{jet}}\right\vert<2.0$~\cite{CMS:2021vui} (triangles) 
are shown with statistical errors (bars) and systematic uncertainties (bands). 
}
\label{fig:atlas_rg_afo_pt_pbpb}
\end{figure*}

\begin{figure*}[htb!]
\centering
\includegraphics[width=\textwidth]{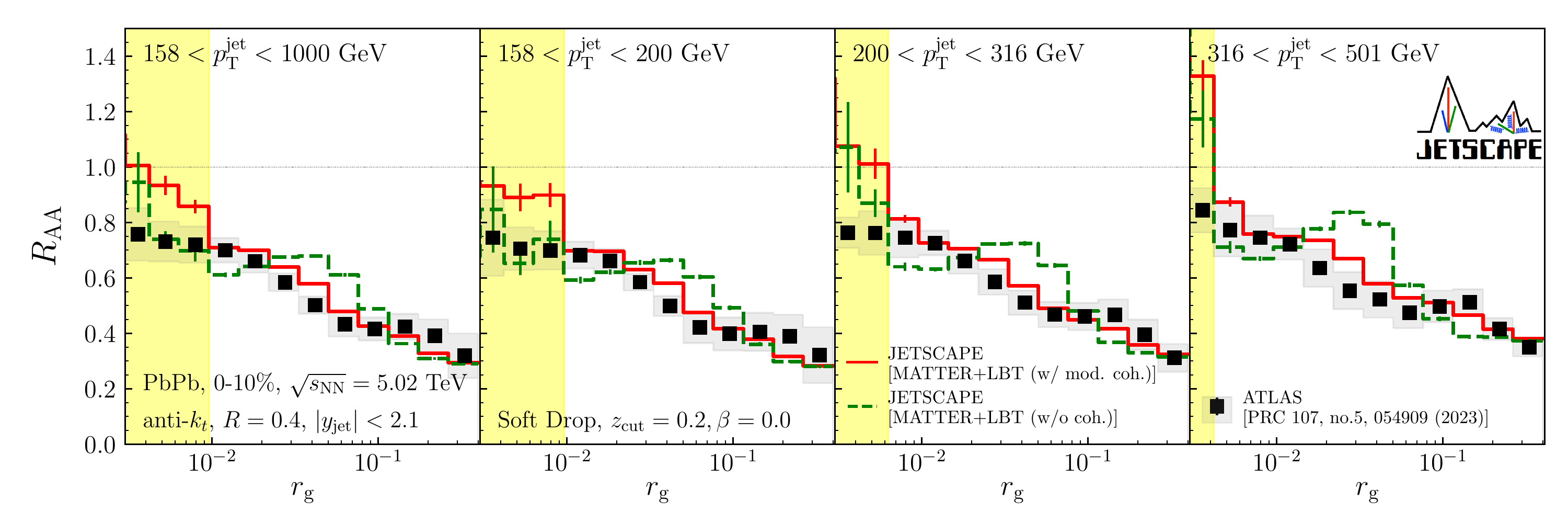}
\caption{(Color online) Nuclear modification factor $R_{AA}$ as a function of $r_{g}$ for jets with different $p_{T}^{\mathrm{jet}}$ in $0\%$--$10\%$ Pb$+$Pb collisions at $\sqrt{s_{\mathrm{NN}}}=5.02$~TeV. 
Jets are reconstructed with $R=0.4$ at midrapidity $|y_{\mathrm{jet}}|<2.1$. 
The Soft Drop parameters are $z_{\mathrm{cut}}=0.2$ and $\beta = 0$. 
The solid and dashed lines with statistical error bars show the results from \textsc{matter}+\textsc{lbt} of \textsc{jetscape} with modified coherence effects and without any coherence effects, respectively. 
For comparison, the experimental data
from the ATLAS Collaboration~\cite{ATLAS:2022vii} are shown by squares with statistical errors (bars) and systematic uncertainties (bands). 
The yellow-shaded regions are the bin areas, including the regime where the perturbation approach does not apply (see text for details). 
}
\label{fig:atlas_rg_afo_rg_pbpb}
\end{figure*}

Motivated by the recent analysis by ATLAS~\cite{ATLAS:2022vii}, 
we also calculated the nuclear modification factor $R_{AA}$ for full jets with different $r_{g}$. 
Figures~\ref{fig:atlas_rg_afo_pt_pbpb} and~\ref{fig:atlas_rg_afo_rg_pbpb} show the $R_{AA}$ results as a function of $p_{T}^{\mathrm{jet}}$ and $r_{g}$, respectively. 
Here, $R_{AA}$ is defined as 
\begin{align}
R_{AA}
=
\frac{
\left.\frac{1}{\langle N_{\mathrm{coll}} \rangle}\frac{d^2 \sigma_{\mathrm{SD,jet}}}{dr_{g} dp_{T}^{\mathrm{jet}}}\right|_{AA}
}{
\left.\frac{d^2 \sigma_{\mathrm{SD,jet}}}{dr_{g} dp_{T}^{\mathrm{jet}}}\right|_{pp}
},
\end{align}
for jets passing the Soft Drop condition with a finite value of $r_g$, and 
\begin{align}
R_{AA}
=
\frac{
\left.\frac{1}{\langle N_{\mathrm{coll}} \rangle}\frac{d \sigma^{\mathrm{incl/}r_{\!g}=0}_{\mathrm{jet}}}{ dp_{T}^{\mathrm{jet}}}\right|_{AA}
}{
\left.\frac{d \sigma^{\mathrm{incl/}r_{\!g}=0}_{\mathrm{jet}}}{ dp_{T}^{\mathrm{jet}}}\right|_{pp}
},
\end{align}
for inclusive jets and jets failing the Soft Drop
condition ($r_{g}=0$), where $\sigma^{\mathrm{incl/}r_{\!g}=0}_{\mathrm{jet}}$ is the cross section of jets for each condition. 
The denominator is calculated for $p$+$p$ collisions, and the numerator is for a given centrality class of $A+A$ collisions, where $\langle N_{\mathrm{coll}} \rangle$ is the average number of binary nucleon-nucleon collisions in the given centrality class. 

\pagelabel{comment4}
Figure~\ref{fig:atlas_rg_afo_pt_pbpb} shows jet $R_{AA}$ as a function of $p_{T}^{\mathrm{jet}}$ for different $r_g$ intervals. 
As already described in Ref.~\cite{JETSCAPE:2022jer}, for the case of inclusive jets (top left plot in Fig.~\ref{fig:atlas_rg_afo_pt_pbpb}), no clear differences due to modified coherence effects are observed in the jet $R_{AA}$. 
Note that the overall medium coupling parameter $\alpha^{\mathrm{fix}}_{s}$ is adjusted separately for each setup ($\alpha^{\mathrm{fix}}_{s}=0.3$ for the case with modified coherence effects, and $0.25$ for the case without coherence effects). 
It should also be noted that our full simulations, which do not incorporate nuclear parton distribution function (NPDF) for nuclear shadowing effects, show a sharper increase than the ATLAS data and slightly larger than the CMS data. Upon incorporation of the NPDF effects, as demonstrated in preceding studies such as Refs.~\cite{Pablos:2019ngg,Caucal:2020uic}, an additional suppression of the $R_{AA}$ for $p_{T} > 300$~GeV, which could consequently align the results more closely with the experimental data, is anticipated. We intend to conduct future research within this more realistic setup systematically.

Moving to the case of Soft Drop fail (top middle plot in Fig.~\ref{fig:atlas_rg_afo_pt_pbpb}), one notices that the data clearly prefer the simulation with modified coherence as opposed to that without coherence. The reduced suppression for the case with modified coherence can be understood under the assumption that the prong structure is established in the high virtuality or \textsc{matter} stage. In this stage, the effective jet quenching strength with the virtuality dependence $\hat{q}_{\mathrm{eff}} \equiv \hat{q}_{\mathrm{HTL}} f(Q^2)$ is smaller for the case with modified coherence effects compared with that without. For the case without coherence, the larger value of $\hat{q}_{\mathrm{eff}}$ in the \textsc{matter} stage leads to the formation of wider prongs, leading to a reduction in the number of jets that fail the Soft Drop condition. 

It bears repeating yet again: The comparisons of simulations to data presented in this paper do \emph{not} include any parameter tuning to fit any substructure data. All parameter tuning was carried out in the calculation of the single high-$p_{T}$ particle and jet suppressions in Ref.~\cite{JETSCAPE:2022jer}. All simulation results presented in this paper are predictions. 

Figure~\ref{fig:atlas_rg_afo_rg_pbpb} shows jet $R_{AA}$ as a function of $r_g$ for different $p_{T}^{\mathrm{jet}}$ intervals. 
The yellow-shaded regions in the figure indicate the areas of bins containing contributions from jets with a transverse scale $\mu_\perp\approxeq p_{T}^{\mathrm{jet}}r_{g} \lessapprox  1~\mathrm{GeV}$, where the perturbative description of parton splitting in the model is not valid. 
To regulate the infra-red singularity in the splitting function, the model needs to specify a minimum cutoff scale for resolvable splitting~\cite{Ellis:1996mzs}, which here is $Q_{\mathrm{min}}=1~\mathrm{GeV}$. 

\pagelabel{revision-comment3-nonpert}
In other words, the jet structure of the yellow-shaded region is governed by the effects from nonperturbative dynamics, namely hydrodynamic evolution of the soft thermalized portion of jets (not modeled in this study), hadronization and subsequent dynamics, rather than the perturbative parton-level dynamics. 
Note that one needs to examine the results shown in Figs.~\ref{fig:alice_rg_pbpb} and~\ref{fig:atlas_rg_afo_pt_pbpb} with the same considerations for regions with small $r_g$ or small $p_{T}^{\mathrm{jet}}$. In this regard, it should also be noted that the results in Fig.~\ref{fig:alice_rg_pbpb} are for charged jets. 

\pagelabel{comment3}
The enhancement or sharp rise of $R_{AA}$ in the nonperturbatively low $r_g$ region, indicated by the yellow band in Fig.~\ref{fig:atlas_rg_afo_rg_pbpb}, is primarily attributed to the cutoff for splittings with the scale $Q_{\mathrm{min}}$. For a parent parton with energy $E$, the minimum splitting angle, below which partonic splitting is vetoed, is approximated as $\theta_{\mathrm{min}} \approxeq Q_{\mathrm{min}}/E$. Given that the value of $Q_{\mathrm{min}} = 1$~GeV is fixed in the model, $\theta_{\mathrm{min}}$ is smaller for jets with a larger initial energy of a parent parton. 

In the presence of jet energy loss due to medium effects, jets with larger initial energies are triggered compared with the vacuum case. As a result, jets are triggered even in the angular region prohibited for the vacuum case, directly leading to enhancement in that region at the parton level. In reality, there is a nonperturbative process involved in hadronization, even in our model calculations. Therefore, this enhancement is slightly blurred, particularly for the region where $p^{\mathrm{jet}}_{\mathrm{T}}$ is not sufficiently large enough to preserve the parton jet substructure. Furthermore, in such a small angle region, where the virtuality is sufficiently small, the case with modified coherence effects a more frequent occurrence of feed-down into the vetoed region due to the larger value of $\alpha^{\mathrm{fix}}_{s}$. Thus, the solid red lines ($R_{AA}$ with modified coherence) always slightly exceed the dashed green lines ($R_{AA}$ calculated without coherence) in the yellow-shaded region.

At very large $r_g$, with $r_g > 0.2$, the prong structure as the transverse scale of the split exceeds $\mu_\perp \gtrapprox 158\,{\rm GeV} \times 0.2 \approx 32$~GeV can be completely dominated by the virtuality acquired by a parent parton at its production in the initial hard scattering. 
This is because, in this region, the initial virtuality is quite large, and furthermore, the formation time for the splitting is very short: $\tau_{\mathrm{form}}\lessapprox \frac{2\cdot (158~\mathrm{GeV})}{(32~\mathrm{GeV})^2} \approx 0.3~\mathrm{GeV}^{-1}\approx 0.06~\mathrm{fm}$. 
Thus, even without the interaction reduction due to modified coherence, no amount of scattering from the medium has much of an effect on the hard splitting. 
As a result, the $R_{AA}$ as a function of $p_{T}^{\mathrm{jet}}$ for the case of $0.2642 < r_g < 0.4$ shows no difference between the cases with modified coherence and without coherence, as shown in the bottom panel of Fig.~\ref{fig:atlas_rg_afo_pt_pbpb}. This is also the case for $r_g \gtrapprox 0.2$ in all the plots of Fig.~\ref{fig:atlas_rg_afo_rg_pbpb}.

\pagelabel{rerevision-comment3-angle}
We finally address the region with $0.022 < r_g < 0.26$. Perturbative QCD should be applicable in this region. Calculations without coherence effects (dashed green line) include a $\hat{q}_{\mathrm{eff}} = \hat{q}_{\mathrm{HTL}} $, that has a large value, even in the high virtuality \textsc{matter} stage, given by Eq.~\eqref{eq:HTL-qhat-formula-C-2}. In this case, multiple scatterings before the hard split, in the high virtuality stage, that leads to the two prongs, provide additional contributions that significantly broaden the prong structure, leading to depletion at lower $r_g$ and an enhancement in the range $0.02 \lessapprox r_g \lessapprox 0.06$, which gradually diminishes at large $r_g \gtrapprox 0.1$. 

To estimate this excess momentum broadening, we consider the relation between the virtuality of the parent hard parton, and the transverse momentum of the two daughter partons. If the parent parton with momentum $p \equiv [p^+,Q^2/(2p^+),0_\perp]$ decays into two daughter partons with $p_1 \equiv [ zp^+, k_\perp^2/(2zp^+), \Vec{k}_\perp ]$ and $p_2 \equiv [ (1-z)p^+, k_\perp^2/\{ 2(1-z)p^+\} , - \Vec{k}_\perp  ] $, we obtain the simple relation that, 
%
\begin{align}
    Q^2 = \frac{k_\perp^2}{  z (1-z)}.
\label{eq:mu-kperp}
\end{align}
To be clear, the $k_\perp$ above is not the final transverse momentum of the prongs. It is one contribution to the total transverse momentum of each prong. 
The maximum value of this contribution takes place when the split has a transverse momentum comparable to the medium generated scale: 
\begin{align}
    k_\perp^2 \approxeq \hat{q} \tau_f,  
    \label{eq:kperp-basic}
\end{align}
where $\tau_f = 2E/Q^2$ is the formation time, of the two hard partons that seed the prongs. We remind the reader that $k_\perp$ in the equation above is not the transverse momentum exchanged with the medium, but the transverse momentum of the radiated partons that will seed the prongs, even though the formula above is typical of the multiple-scattering low virtuality stage. The reason that scatterings before the hard split have their largest contribution at $k_\perp^2 \simeq \hat{q} \tau$ is because the medium modified portion of the splitting function in Eq.~\eqref{eq:HT-splitting function} reaches its largest value at this point. 

Including all terms together, we obtain, 
\begin{align}
    k_\perp^2 \approxeq \sqrt{ z(1-z) 2 \hat{q} E }.
    \label{eq:borad_esti}
\end{align}
%
Taking the simple case that $z=1-z = 1/2$, we obtain, $k_\perp^2 = \sqrt{\hat{q}E/4}$. This yields the simple expression for the peak angle of the bump of the dashed green line as, 
%
\begin{align}
    \theta_{\mathrm{max}} &\approxeq \theta^1_{\mathrm{max}} \!\!+ \theta^2_{\mathrm{max}} =  \frac{k_\perp}{zE}\! +\! \frac{k_\perp}{(1-z)E} \approx \frac{ 2\sqrt{2}(\hat{q})^{\frac{1}{4}} }{E^{\frac{3}{4}}}.
    \label{eq:theta_max}
\end{align}
%
In the above equation, the $\hat{q}$ would vary with the path taken by the hard parton, and thus, its appearance in the expression above is meant to indicate an approximate,  averaged value of $\hat{q}$. As such, we cannot use the above equation to deduce the exact value of $\theta_{\mathrm{max}}$. However, one can deduce the approximate dependence of $\theta_{\mathrm{max}}$ on the energy $E$.

Using the above equation, one would obtain that if the energy of the jet were to double, the angle of the bump in the dashed green line in Fig.~\ref{fig:atlas_rg_afo_rg_pbpb} would move down in $r_g$ by a factor of $2^{3/4} \approx 1.6$. One notes that this is indeed the case in the $2^{\rm nd}$ and $4^{\rm th}$ panels of Fig.~\ref{fig:atlas_rg_afo_rg_pbpb}. The energy range between these doubles, and the position of the bump in the green curve shifts down in $r_g$ by about a factor of $1.5$-$2$. 
This different behavior, depending on the presence or absence of modified coherence effects, is also  evident when shown as a function of $p_{T}^{\mathrm{jet}}$ from intermediate ranges of $r_g$, as in Fig.~\ref{fig:atlas_rg_afo_pt_pbpb}. 

\pagelabel{comment2}
In contrast, when modified coherence effects are present (solid red line), a clear bump structure is not readily apparent. The distortion of the $r_{g}$ distributions driven by the small momentum broadening, significantly suppressed by modified coherence, can occur at very small $r_{g}$. In these small $r_{g}$ regions, another enhancement attributed to jet energy loss, as discussed earlier, comes into play and makes the structure of the modification, possibly a bump, due to the broadening less noticeable.

In the modification of the jet $r_{g}$ distribution, multiple factors contribute and compete with each other. Moreover, in practice, these factors originate from the same jet-medium interaction, making it impossible, even in theoretical simulations, to eliminate just one of them. Thus, conducting an analysis to distinguish between these multiple contributing factors is quite challenging with the event setup used in this study. However, it might be feasible to a certain extent, for example, by using photon-tagged jet events. By triggering jets using the $p_{T}$ of photons that tag them--thereby inheriting the initial $p_{T}^{\mathrm{jet}}$--we can have better control over the energy-loss effects on jets. We leave this in-depth study for future work. 

\pagelabel{rerevision-comment3-bump}
The bump structure in the jet $R_{AA}$, as a function of $r_g$, which our results without modified coherence effects show, can also be seen in the results from 
other  models (with different assumptions and implementations)~\cite{Caucal:2019uvr,Caucal:2020uic,Caucal:2018dla,Ringer:2019rfk}. In contrast, the data from ATLAS exhibit an almost monotonically decreasing trend with no such clear bump structure for all $p_{T}^{\mathrm{jet}}$ intervals. This agrees with the \textsc{matter}+\textsc{lbt} result with modified coherence. This reveals that the medium effect for the hard splitting modification is strongly suppressed. 

\begin{figure*}[htbp]
\centering
\includegraphics[width=0.75\textwidth]{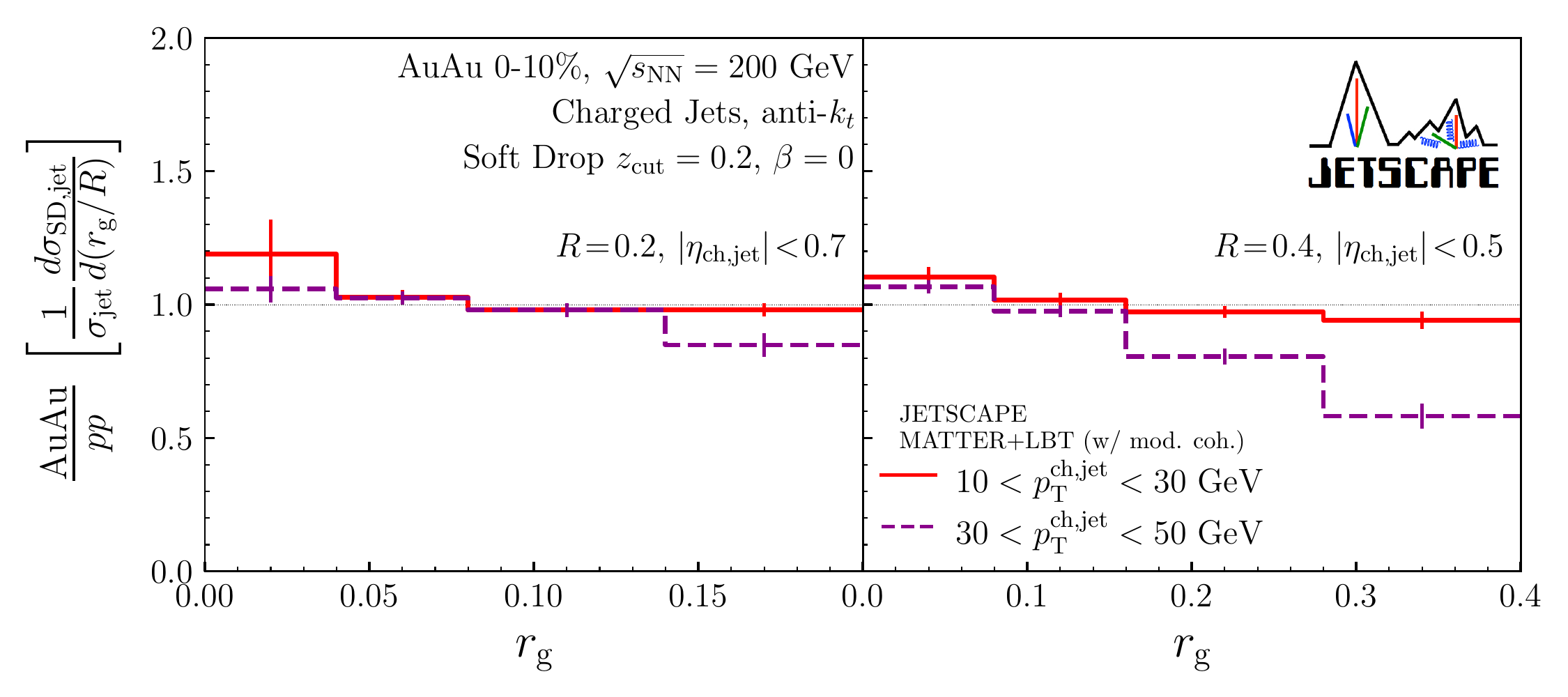}
\caption{(Color online) 
Ratios of $r_{g}$ distributions for charged jets with $R=0.2$ and $|\eta_{\mathrm{ch,jet}}|<0.7$ (left), and $R=0.4$, $|\eta_{\mathrm{ch,jet}}|<0.5$ (right) between $0\%$--$10\%$ Au$+$Au and $p$+$p$ collisions at $\sqrt{s_{\mathrm{NN}}}=200$~GeV, from \textsc{matter}+\textsc{lbt} simulations within \textsc{jetscape}, including modified coherence effects. 
The Soft Drop parameters are $z_{\mathrm{cut}}=0.2$ and $\beta = 0$. 
The solid and dashed lines with statistical error bars show the results for $10<p^{\mathrm{ch,jet}}_{T}<30$~GeV and $30<p^{\mathrm{ch,jet}}_{T}<50$~GeV, respectively. 
}
\label{fig:rhic_rg}
\end{figure*}

Figure~\ref{fig:rhic_rg} presents our prediction for the modification of $r_{g}$ distribution for charged jets in $0\%$--$10\%$ Au$+$Au collisions at $\sqrt{s_{\mathrm{NN}}}=200$~GeV from \textsc{matter}+\textsc{lbt} with modified coherence effects. 
Similar to the LHC case, one finds enhancement at small $r_g$ and slight suppression at large $r_g$, which is more pronounced for jets with larger transverse momentum. 

\subsection{Jet fragmentation function} 

\begin{figure*}[htb!]
\centering
\includegraphics[width=0.98\textwidth]{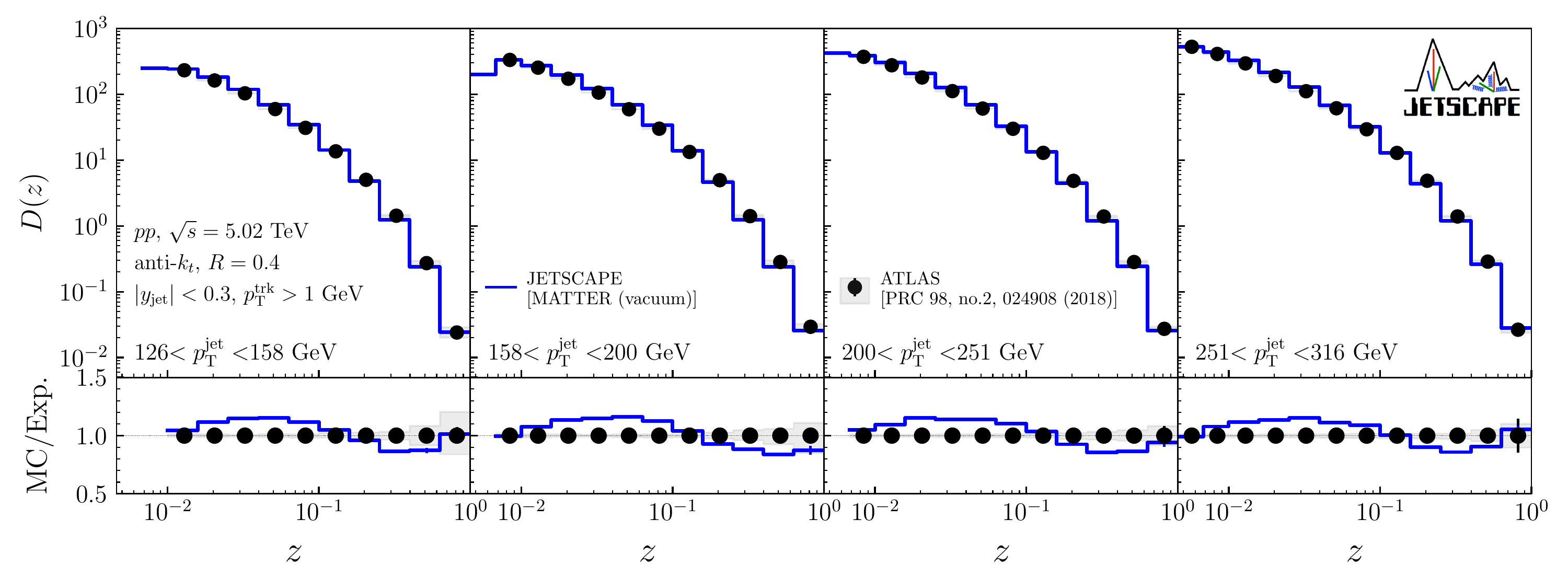}
\includegraphics[width=0.98\textwidth]{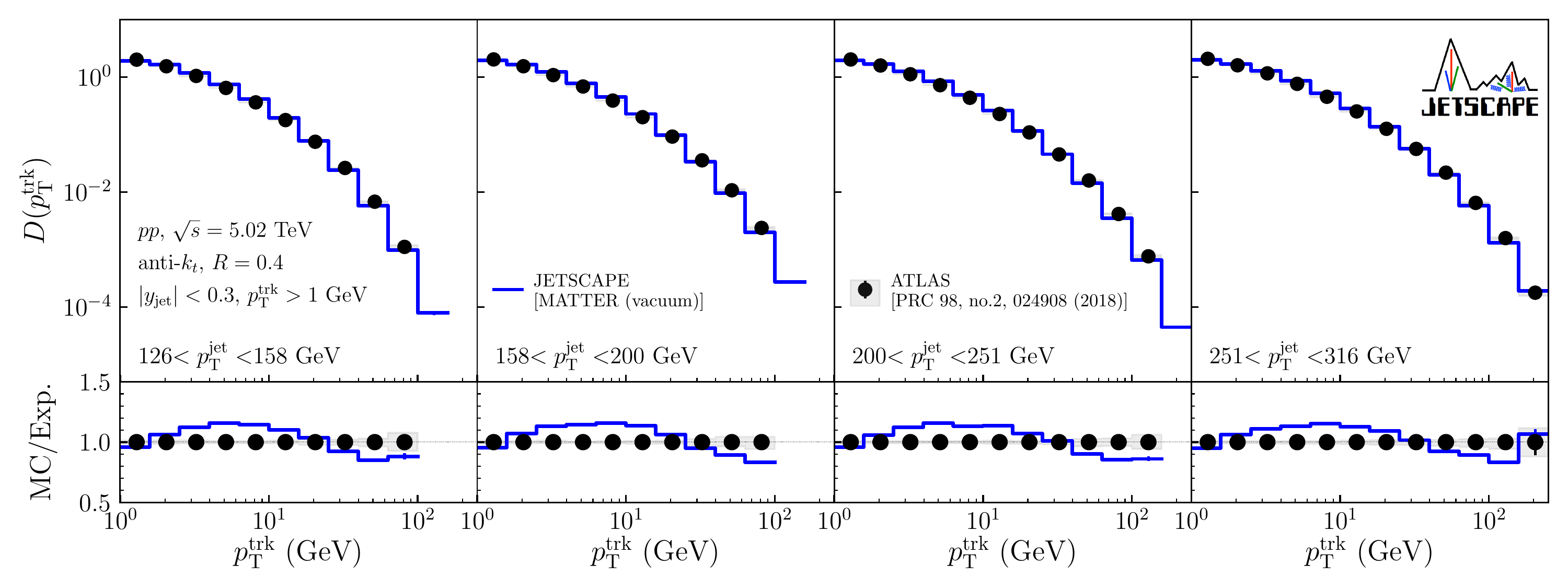}
\caption{(Color online) Jet fragmentation functions for jets in $p$+$p$ collisions at $\sqrt{s}=5.02$~TeV and the ratios as a function of $z$ (top) and $p^{\mathrm{trk}}_{T}$ (bottom) for different $p_{T}^{\mathrm{jet}}$ range. 
Jets are fully reconstructed including both charged and neutral particles by anti-$k_{t}$ with $R=0.4$ at midrapidity $\left|y^{\mathrm{jet}}\right|<0.3$. 
The solid lines and circles with statistical error bars show the results from \textsc{jetscape} and the experimental data from the ATLAS Collaboration~\cite{ATLAS:2018bvp}, respectively. 
The bands indicate the systematic uncertainties of the experimental data. }
\label{fig:atlas_ff_pp}
\end{figure*}

We now turn to the last jet substructure observable: the jet fragmentation function. 
Jet fragmentation functions are measured as a function of 
the track-particle transverse momentum $p^{\mathrm{trk}}_{T}$
or longitudinal momentum fraction relative to the jet, 
\begin{align}
z&=\frac{p^{\mathrm{trk}}_{T}\cos(\Delta r)}{p_{T}^{\mathrm{jet}}}, 
\end{align}
where $\Delta r = [(\eta_{\mathrm{trk}}-\eta_{\mathrm{jet}})^2+(\phi_{\mathrm{trk}}-\phi_{\mathrm{jet}})^2]^{1/2}$. 
The fragmentation functions are defined as 
\begin{align}	
D(z)&=\frac{1}{N_{\mathrm{jet}}}\frac{dN_\mathrm{trk}}{dz},\\
D(p^{\mathrm{trk}}_{T})&=\frac{1}{N_{\mathrm{jet}}}\frac{dN_\mathrm{trk}}{dp^{\mathrm{trk}}_{T}},
\end{align}
where $N_{\mathrm{jet}}$ is the number of triggered jets 
and $N_{\mathrm{trk}}$ is the number of charged track particles detected inside the jet cones, $\Delta r < R$. 
Our \textsc{jetscape} PP19 results for the fragmentation functions are compared with the experimental data by ATLAS in Fig.~\ref{fig:atlas_ff_pp}. 
For all available $p_{T}^{\mathrm{jet}}$ ranges, the discrepancies from the data are generally within $20\%$ at most.

\begin{figure*}[htb!]
\centering
\includegraphics[width=0.98\textwidth]{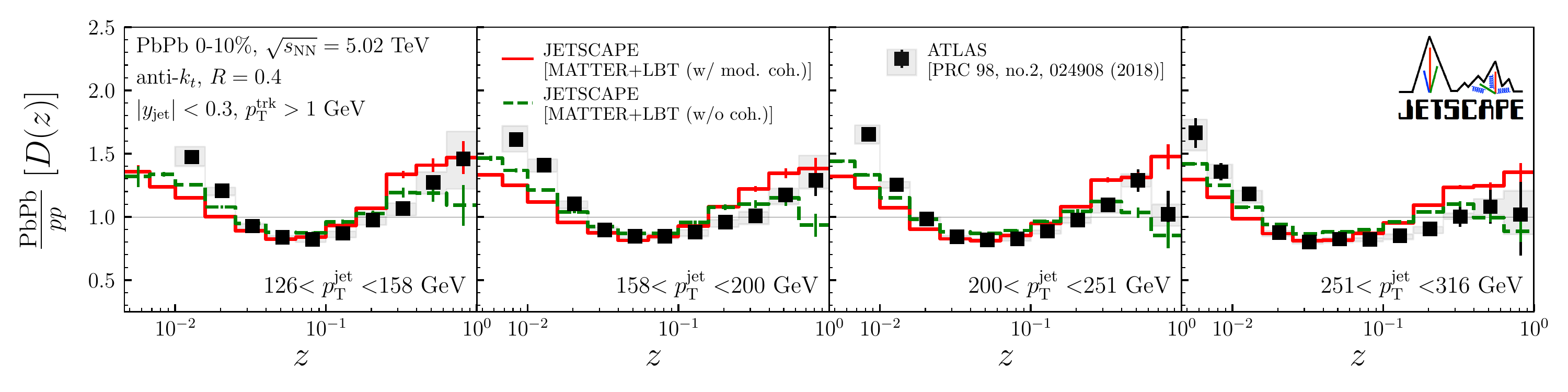}
\includegraphics[width=0.98\textwidth]{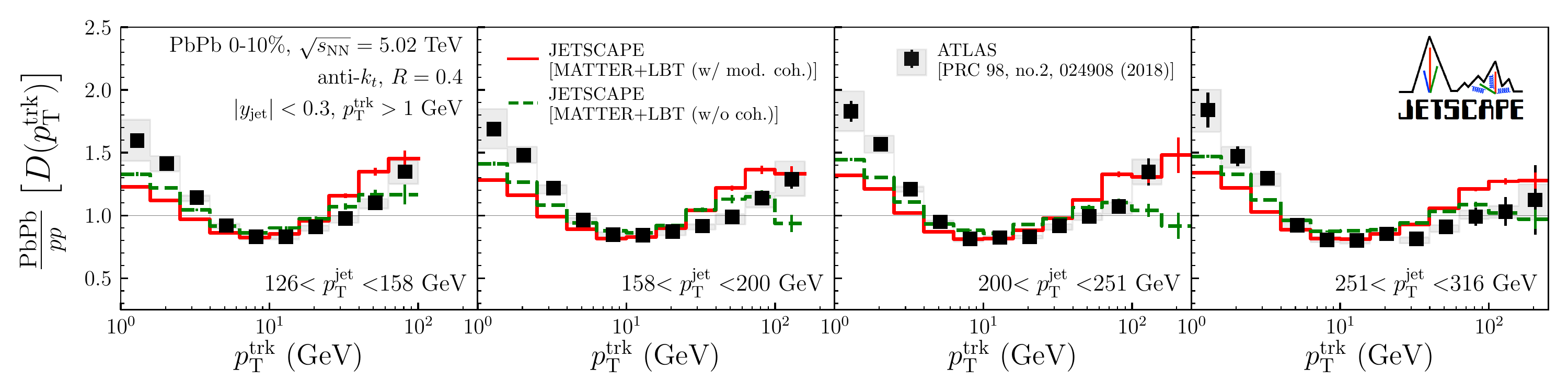}
\caption{(Color online) 
Ratios of jet fragmentation functions for jets 
between $0\%$--$10\%$ Pb$+$Pb and $p$+$p$ collisions at $\sqrt{s_{\mathrm{NN}}}=5.02$~TeV 
as a function of $z$ (top) and $p^{\mathrm{trk}}_{T}$ (bottom) for different $p_{T}^{\mathrm{jet}}$ range. 
Jets are fully reconstructed, including both charged and neutral particles by anti-$k_{t}$ with $R=0.4$ at midrapidity $\left|y^{\mathrm{jet}}\right|<0.3$. 
The solid and dashed with statistical error bars lines show the results from \textsc{matter}+\textsc{lbt} of \textsc{jetscape} with modified coherence effects and without any coherence effects, respectively. 
For comparison, the experimental data from the ATLAS Collaboration~\cite{ATLAS:2018bvp} are shown by squares with statistical errors (bars) and systematic uncertainties (bands). 
}
\label{fig:atlas_ff_pbpb}
\end{figure*}

In Fig.~\ref{fig:atlas_ff_pbpb}, we present the modification of the jet fragmentation functions for full jets in $0\%$--$10\%$ Pb$+$Pb collisions at $\sqrt{s_{\mathrm{NN}}}=5.02$~TeV. 
Results from the \textsc{matter}+\textsc{lbt} simulations, both with modified coherence effects and without any coherence effects, are compared with the experimental data from ATLAS. 
All the simulation results and the data show qualitatively the same trends. 
While the track particles at intermediate $z$ are suppressed by the interactions with the medium and give the enhancement at small $z$, the large-$z$ part is enhanced due to the less affected hard part of jets. 

In jet fragmentation functions, modified coherence effects are quantitatively visible as more prominent enhancements in the large-$z$ region dominated by hadrons from leading partons of jets. 
Since the leading parton has the largest virtuality at the early stage in the jet shower evolution, the interaction reduction due to modified coherence affects this parton the most. 
As a result, the modification of large-$z$ jet hadrons is further lessened, and the enhancement becomes more substantial than the case without coherence effects. 
This is consistent with the weak energy loss of inclusive charged particles at high $p_{T}$ explained by modified coherence effects presented in Refs.~\cite{JETSCAPE:2022jer,JETSCAPE:2022hcb,JETSCAPE:2023ikg}. 

\pagelabel{rerevision-comment4-fragf}
Both results with modified coherence effects and without any coherence effects show a sizable enhancement at low-$z$ mainly due to the medium response via recoils but still underestimate the data. In the presence of the modified coherence effect, the enhancement in the low-$z$ region is slightly less prominent than in the absence of the modified coherence effect, in conjunction with the stronger enhancement in the high-$z$ region.
For the underestimation at low-$z$, one possible cause is the visible discrepancy in the suppression at mid-$z$. 
Furthermore, for some very soft components of jets giving contribution in the low-$z$ region, the recoil prescription may not provide an entirely reasonable description once their energies become close to the typical scale for the medium constituents. 
More comprehensive momentum structures of jet constituents, including such soft regions where hydrodynamic medium response needs to be considered, will be explored in a future effort. 

With the current uncertainties, it is not yet possible to conclude the presence of modified coherence effects from comparisons with only the experimental data on modified jet fragmentation functions. However, when taken in conjunction with the results on the $r_g$ distribution, a stronger case can be made for the existence of modified coherence effects at high virtuality. 
Our results also indicate that the medium effects over different scales can be discernible by future measurements with high precision.

\begin{figure*}[htbp]
\centering
\includegraphics[width=0.75\textwidth]{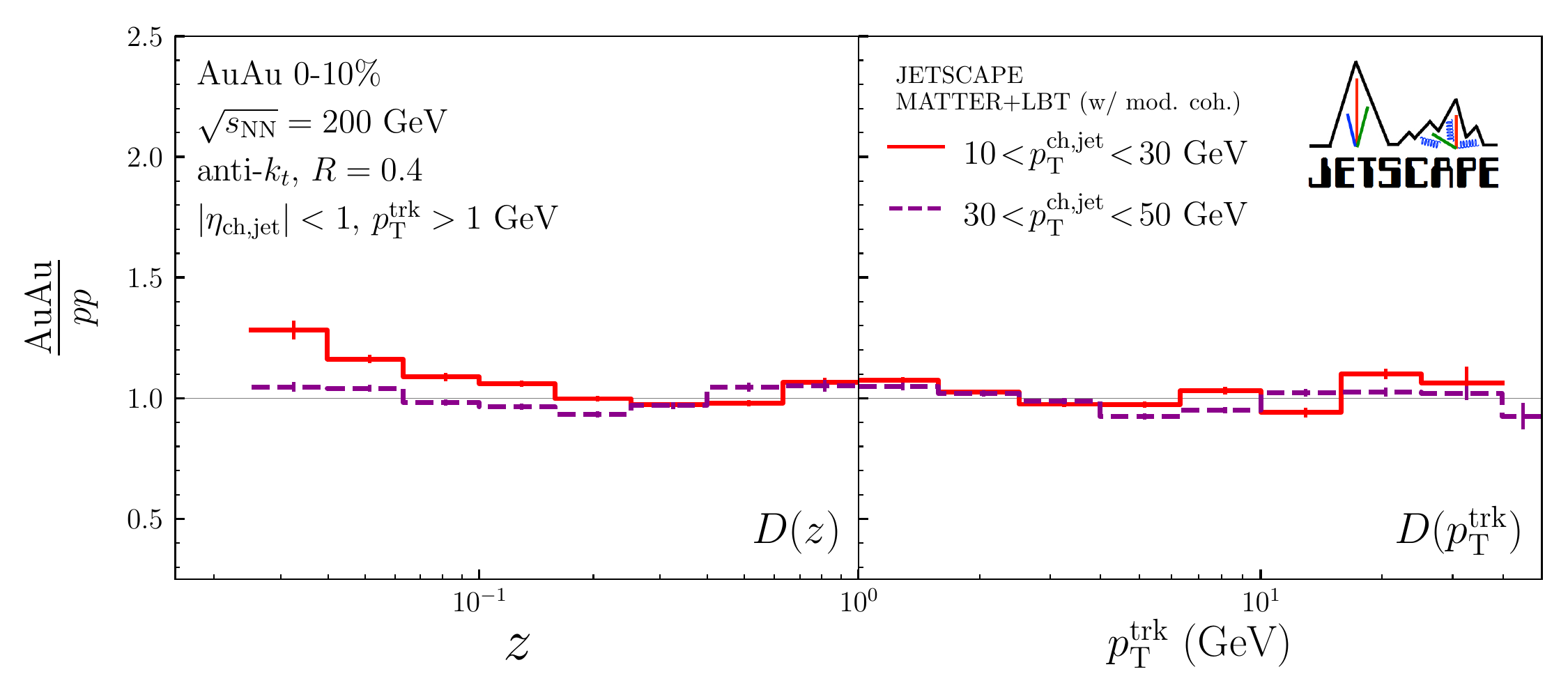}
\caption{(Color online) 
Ratios of jet fragmentation functions for charged jets with $R=0.4$ and $\left|\eta_{\mathrm{ch, jet}}\right|<1.0$ between $0\%$--$10\%$ Au$+$Au and $p$+$p$ collisions at $\sqrt{s_{\mathrm{NN}}}=200$~GeV 
as a function of $z$ (left) and $p^{\mathrm{trk}}_{T}$ (right) 
from \textsc{matter}+\textsc{lbt} of \textsc{jetscape} with modified coherence effects. 
The solid and dashed lines with statistical error bars show the results for $10<p^{\mathrm{ch,jet}}_{T}<30$~GeV and $30<p^{\mathrm{ch,jet}}_{T}<50$~GeV, respectively.
}
\label{fig:rhic_ff}
\end{figure*}

In Fig.~\ref{fig:rhic_ff}, we present our results of the modification of jet fragmentation functions for charged jets in $0\%$--$10\%$ Au$+$Au collisions at $\sqrt{s_{\mathrm{NN}}}=200$~GeV from \textsc{matter}+\textsc{lbt} with modified coherence effects. 
Compared with the results for the top LHC energy, the modifications are quite small. 

\section{Summary and Outlook}
\label{Section:Summary}
This paper explored the medium modification of jet substructure in high-energy heavy-ion collisions, employing a multistage jet evolution model, \textsc{matter}+\textsc{lbt}, with the configuration and parameters established within the \textsc{jetscape} framework by comparison with leading hadron and jet data. All parameters were taken from our previous efforts~\cite{JETSCAPE:2022jer} and were not retuned for this study. In fact, no new simulations were run for this paper. The presented results were calculated from the simulations carried out for Ref.~\cite{JETSCAPE:2022jer}. 

To investigate the contribution of modified coherence effects based on the ability of the medium to resolve the partons radiated from splits at high energy and virtuality, we performed numerical simulations for two cases, with modified coherence effects and without any coherence effects. 
These modified coherence effects are implemented as the $Q^2$-dependent modulation factor in the medium-modified splitting function and give a drastic reduction of the interaction with the medium with increasing parton virtuality. 

The distribution of jet splitting momentum fraction ($z_{g}$) shows almost no visible modification due to the medium effects for any kinematic configurations in both cases with modified coherence effects and without coherence effects. This extremely small sensitivity to the medium effects is consistent with the experimental data taken by ALICE at the LHC. Our predictions for future RHIC measurements also show no significant modification. 

Then, we presented the observables related to the jet splitting radius ($r_{g}$). 
In comparison with the ALICE data, both results with modified coherence effects and without coherence effects satisfactorily capture the monotonically decreasing behavior with increasing radius and give good agreement. Here, no conclusions about coherent effects could be drawn from this analysis in comparison with the data from ALICE. We reiterate again that our simulations reduce to and reproduce the $z_g$ and $r_g$ distributions in the absence of the medium, in comparison with data from $p$+$p$ collisions. 

In comparison with data from ATLAS~\cite{ATLAS:2022vii}, we demonstrated that modified coherence effects manifest, even at the qualitative behavior level, in $r_{g}$-dependent $R_{AA}$ with finer binning. In both the $R_{AA}$ as a function of $p_{T}^{\mathrm{jet}}$ for different bins of the angle $r_g$ as well as the $R_{AA}$ as a function of $r_g$ in different $p_{T}^{\mathrm{jet}}$ bins, there is a clear difference between simulations with modified coherence and without coherence. The experimental data clearly prefer simulations with modified coherence effects. This indicates that the scattering with the medium constituents at high virtuality is reduced due to the finer scale of the medium probed by the jet parton. 

Finally, we found that modified coherence effects may also be visible as a more prominent enhancement at large $z$ in the modification pattern of the jet fragmentation functions. 
The energy loss of hard leading partons, which form the jet core components with large transverse momentum, is highly suppressed by modified coherence effects due to their large virtualities. The data have a slight preference for simulations with modified coherence if one restricts attention to particles with $z\gtrapprox 0.1$. For both cases with modified coherence and without coherence, the simulations produce fewer particles at very small $z$ ($z \lessapprox 0.02$), with the case without coherence performing marginally better. 

This paper constitutes the third installment of jet and hadron-based observables from the \textsc{matter}+\textsc{lbt} simulations in the \textsc{jetscape} framework~\cite{JETSCAPE:2022jer,JETSCAPE:2022hcb,JETSCAPE:2023ikg}. In all three of these papers, including the current effort, we have demonstrated wide-ranging agreement for the hard sector of jets, between simulations, typically with modified coherence and experimental data. The only remaining issues within the hard sector of the jet are related to coincidence measurements. These will be presented in a future effort. 

In terms of physics included within these simulations, the one remaining component is the very soft sector of jets. In the current effort, this was pointed out in the discussion of the low $r_g$ section of the $r_g$ dependent $R_{AA}$, and the low-$z$ and low-$p_{T}$ sector of the jet fragmentation function. 
This requires incorporating an energy deposition scheme in which partons with energy comparable to the ambient temperature are converted into an energy-momentum source term and then included back in the hydrodynamic calculation. 
As may be obvious, these simulations require close to a single hydro run per hard event and, as such, are very computationally demanding. Various schemes to approximately incorporate soft physics without the need for full hydrodynamic simulation are currently underway. The analysis of certain jet-based observables predominantly sensitive to the soft sector of jets will be carried out after these efforts are complete.

\acknowledgments
\label{Ack}
This work was supported in part by the National Science Foundation (NSF) within the framework of the JETSCAPE Collaboration, under grant number OAC-2004571 (CSSI:X-SCAPE). It was also supported under ACI-1550172 (Y.C. and G.R.), ACI-1550221 (R.J.F., F.G., and M.K.), ACI-1550223 (U.H., L.D., and D.L.), ACI-1550225 (S.A.B., T.D., W.F., C.Si., and R.W.), ACI-1550228 (J.M., B.J., P.J., X.-N.W.), and ACI-1550300 (S.C., A.K., J.L., A.M., H.M., C.N., A.S., J.P., L.S., C.Si., I.S., R.A.S., and G.V.); by PHY-1516590 and PHY-1812431 (R.J.F., M.K., and A.S.); it was supported in part by NSF CSSI grant number \rm{OAC-2004601} (BAND; D.L. and U.H.); it was supported in part by the US Department of Energy, Office of Science, Office of Nuclear Physics under grant numbers \rm{DE-AC02-05CH11231} (X.-N.W.), \rm{DE-FG02-00ER41132} (D.O), \rm{DE-AC52-07NA27344} (A.A., R.A.S.), \rm{DE-SC0013460} (S.C., A.K., A.M., C.S., I.S., and C.Si.), \rm{DE-SC0021969} (C.S. and W.Z.), \rm{DE-SC0004286} (L.D., U.H., and D.L.), \rm{DE-SC0012704} (B.S.), \rm{DE-FG02-92ER40713} (J.P.) and \rm{DE-FG02-05ER41367} (S.A.B, T.D., W.F., J.-F.P., C.Si. and D.S.). The work was also supported in part by the National Science Foundation of China (NSFC) under grant numbers 11935007, 11861131009, and 11890714 (Y.H. and X.-N.W.), under grant numbers 12175122 and 2021-867 (S.C.), by the Natural Sciences and Engineering Research Council of Canada (C.G., M.H., S.J., and G.V.), by the University of Regina President's Tri-Agency Grant Support Program (G.V), by the Canada Research Chair program (G.V. and A.K.) reference number CRC-2022-0014,
by the Office of the Vice President for Research (OVPR) at Wayne State University (Y.T.), 
by JSPS KAKENHI Grant No.~22K14041 (Y.T.), by the S\~{a}o Paulo Research Foundation (FAPESP) under projects 2016/24029-6, 2017/05685-2 and 2018/24720\%--6 (A. L. and  M.L.), and by the University of California, Berkeley - Central China Normal University Collaboration Grant (W.K.). U.H. would like to acknowledge support by the Alexander von Humboldt Foundation through a Humboldt Research Award. C.S. acknowledges a DOE Office of Science Early Career Award. Computations were carried out on the Wayne State Grid funded by the Wayne State OVPR. The bulk medium simulations were done using resources provided by the Open Science Grid (OSG) \cite{Pordes:2007zzb, Sfiligoi:2009cct}, which is supported by the National Science Foundation award \#2030508. Data storage was provided in part by the \textsc{osiris} project supported by the National Science Foundation award \#2030508.

\begin{figure*}[htbp]
\centering
\includegraphics[width=0.75\textwidth]{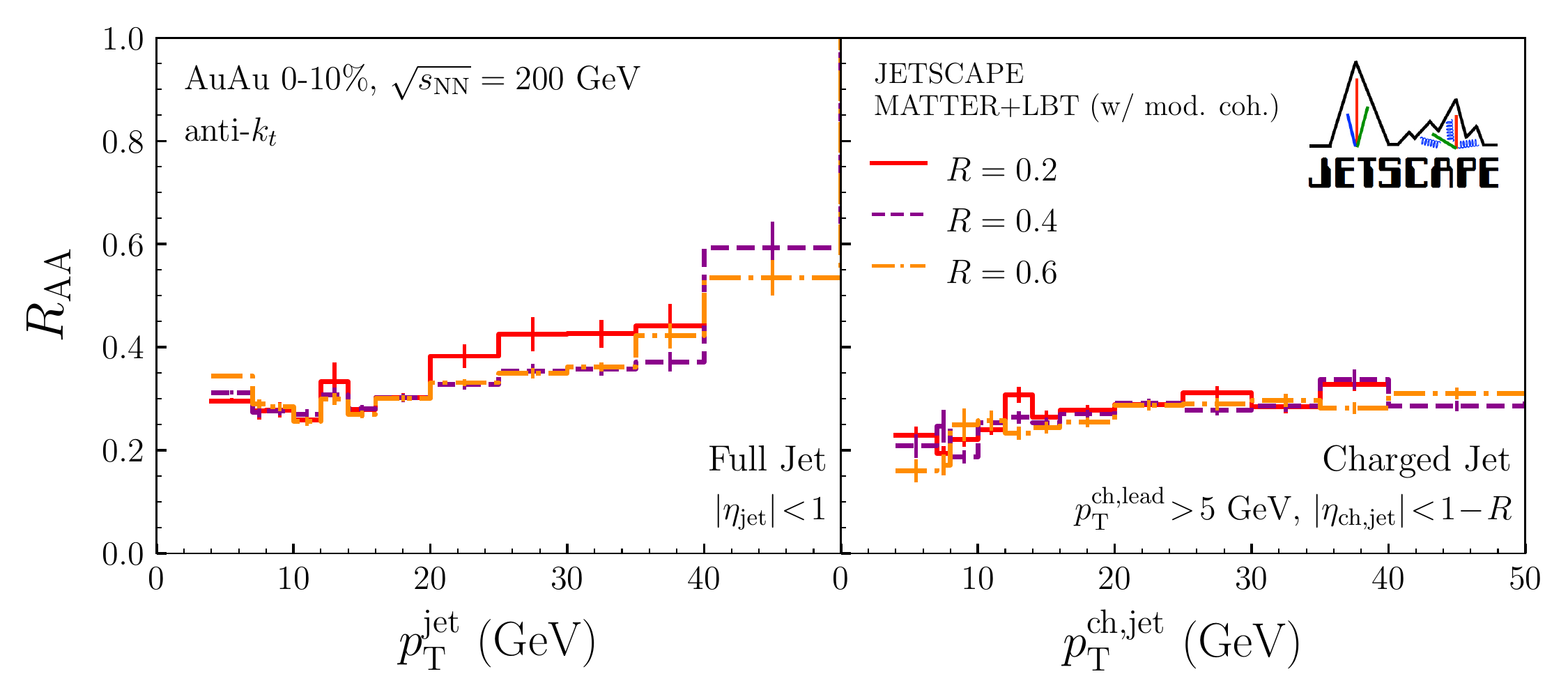}
\caption{(Color online) 
Nuclear modification factor $R_{AA}$ for inclusive full jet with $|\eta_{\mathrm{jet}}|<1$ (left), and charged jet with $|\eta_{\mathrm{ch,jet}}|<1-R$ and leading charged particle $p^{\mathrm{ch,lead}}_{T}>5$ GeV (right) in $0\%$--$10\%$ Au$+$Au collisions at $\sqrt{s_{\mathrm{NN}}}=200$~GeV from \textsc{matter}+\textsc{lbt} of \textsc{jetscape} with modified coherence effects. 
The solid, dashed, and dash-dotted lines with statistical error bars show the results for $R=0.2$, $R=0.4$, and $R=0.6$, respectively. 
}
\label{fig:rhic_raa}
\end{figure*}

\hypertarget{Appen}{}
\appendix*
\section{Jets suppression at RHIC}
For the benchmarking purposes for our jet substructure results in Au$+$Au collisions at $\sqrt{s_{\mathrm{NN}}}=200~\mathrm{GeV}$ presented in the main body of the paper, we also show the predictions of $R_{AA}$ for inclusive full and charged jets from the same event generation by \textsc{matter}+\textsc{lbt} with modified coherence effects in Fig.~\ref{fig:rhic_raa}.

\bibliography{main,manual,misc}

\end{document}